\input harvmac
\input graphicx

\def\Title#1#2{\rightline{#1}\ifx\answ\bigans\nopagenumbers\pageno0\vskip1in
\else\pageno1\vskip.8in\fi \centerline{\titlefont #2}\vskip .5in}
%

%
%
\ifx\includegraphics\UnDeFiNeD\message{(NO graphicx.tex, FIGURES WILL BE IGNORED)}
\def\figin#1{\vskip2in}
\else\message{(FIGURES WILL BE INCLUDED)}\def\figin#1{#1}
\fi
\def\Fig#1{Fig.~\the\figno\xdef#1{Fig.~\the\figno}\global\advance\figno
 by1}
%
%
%
%
\def\Ifig#1#2#3#4{
\goodbreak\midinsert
\figin{\centerline{
\includegraphics[width=#4truein]{#3}}}
\narrower\narrower\noindent{\footnotefont
{\bf #1:}  #2\par}
\endinsert
}
%
%
\font\ticp=cmcsc10
\def\subsubsec#1{\noindent{\undertext {#1}}}
\def\undertext#1{$\underline{\smash{\hbox{#1}}}$}

\def\hf{{1\over 2}}

\def\calo{{\cal O}}
\def\calh{{\cal H}}
\def\caln{{\cal N}}
\def\cals{{\cal S}}

\def\hext{{{\cal H}_{\rm ext}}}
\def\hnear{{{\cal H}_{\rm near}}}
\def\hfar{{{\cal H}_{\rm far}}}
\def\hbh{{{\cal H}_{\rm BH}}}
\def\ha{{{\cal H}_{\rm A}}}
\def\hb{{{\cal H}_{\rm B}}}
\def\hc{{{\cal H}_{\rm C}}}

\def\roughly#1{\mathrel{\raise.3ex\hbox{$#1$\kern-.75em\lower1ex\hbox{$\sim$}}}}

\def\ahat{{\hat a}}
\def\ahats{|\ahat\rangle}
\def\as{|a\rangle}

\def\zhat{{\hat 0}}
\def\ohat{{\hat 1}}
\def\zhats{|\zhat\rangle}
\def\ohats{|\ohat\rangle}
\def\zs{|0\rangle}
\def\os{|1\rangle}
\def\qhat{{\hat q}}
\def\uhat{{\hat U}}
\def\vecm{{\vec m}}
\def\tbh{{T_{BH}}}
\def\mthsu{\mathsurround=0pt  }
\def\leftrightarrowfill{$\mthsu \mathord\leftarrow\mkern-6mu\cleaders
  \hbox{$\mkern-2mu \mathord- \mkern-2mu$}\hfill
  \mkern-6mu\mathord\rightarrow$}
 \def\overleftrightarrow#1{\vbox{\ialign{##\crcr\leftrightarrowfill\crcr\noalign{\kern-1pt\nointerlineskip}$\hfil\displaystyle{#1}\hfil$\crcr}}}
\overfullrule=0pt
%
%
\lref\Hawkrad{
  S.~W.~Hawking,
  ``Particle Creation By Black Holes,''
  Commun.\ Math.\ Phys.\  {\bf 43}, 199 (1975)
  [Erratum-ibid.\  {\bf 46}, 206 (1976)].
}
\lref\Pagerev{
  D.~N.~Page,
  ``Black hole information,''
[hep-th/9305040].
}
\lref\SGTrieste{S.~B.~Giddings,
  ``Quantum mechanics of black holes,''
  arXiv:hep-th/9412138\semi
    ``The black hole information paradox,''
  arXiv:hep-th/9508151.
}
\lref\Mathurrev{
  S.~D.~Mathur,
  ``The Information paradox: A pedagogical introduction,''
Class.\ Quant.\ Grav.\  {\bf 26}, 224001 (2009).
[arXiv:0909.1038 [hep-th]];  ``What the information paradox is {\it not},''  
  [arXiv:1108.0302 [hep-th]].
}
\lref\Haag{
  R.~Haag,
  {\sl Local quantum physics: Fields, particles, algebras,}
Berlin, Germany: Springer (1992) 356 p. (Texts and monographs in physics).
}
\lref\BHMR{
  S.~B.~Giddings,
  ``Black holes and massive remnants,''
Phys.\ Rev.\  {\bf D46}, 1347-1352 (1992).
[hep-th/9203059].
}
\lref\Erice{
  S.~B.~Giddings,
  ``The gravitational S-matrix: Erice lectures,''
[arXiv:1105.2036 [hep-th]].
}
\lref\LQGST{
  S.~B.~Giddings,
  ``Locality in quantum gravity and string theory,''
Phys.\ Rev.\  {\bf D74}, 106006 (2006).
[hep-th/0604072].
}
\lref\NLvC{
  S.~B.~Giddings,
  ``Nonlocality versus complementarity: a conservative approach to the information problem,''
Class.\ Quant.\ Grav.\  {\bf 28}, 025002 (2011).
[arXiv:0911.3395 [hep-th]].
}
\lref\Page{
  D.~N.~Page,
  ``Information in black hole radiation,''
  Phys.\ Rev.\ Lett.\  {\bf 71}, 3743 (1993)
  [arXiv:hep-th/9306083].
}
\lref\HaPr{
  P.~Hayden, J.~Preskill,
  ``Black holes as mirrors: Quantum information in random subsystems,''
JHEP {\bf 0709}, 120 (2007).
[arXiv:0708.4025 [hep-th]].
}
\lref\QBHB{
  S.~B.~Giddings,
  ``Quantization in black hole backgrounds,''
  Phys.\ Rev.\  D {\bf 76}, 064027 (2007)
  [arXiv:hep-th/0703116].
}
\lref\Rozali{
  B.~Czech, K.~Larjo, M.~Rozali,
  ``Black Holes as Rubik's Cubes,''
[arXiv:1106.5229 [hep-th]].
}
\lref\Sussrem{
  L.~Susskind,
  ``Trouble for remnants,''
[hep-th/9501106].
}
\lref\wabhip{
  S.~B.~Giddings,
  ``Why aren't black holes infinitely produced?,''
Phys.\ Rev.\  {\bf D51}, 6860-6869 (1995).
[hep-th/9412159].
}
\lref\SeSu{
  Y.~Sekino, L.~Susskind,
  ``Fast Scramblers,''
JHEP {\bf 0810}, 065 (2008).
[arXiv:0808.2096 [hep-th]].
}
\lref\GiSh{S.~B.~ Giddings and Y.~ Shi, work in progress.}
\lref\fuzz{
  S.~D.~Mathur,
  ``Fuzzballs and the information paradox: A Summary and conjectures,''
[arXiv:0810.4525 [hep-th]].
}

\lref\Hawkunc{
  S.~W.~Hawking,
  ``Breakdown Of Predictability In Gravitational Collapse,''
  Phys.\ Rev.\  D {\bf 14}, 2460 (1976).
}
\lref\Astrorev{
  A.~Strominger,
  ``Les Houches lectures on black holes,''
  arXiv:hep-th/9501071.
}
\lref\LPSTU{
  D.~A.~Lowe, J.~Polchinski, L.~Susskind, L.~Thorlacius and J.~Uglum,
  ``Black hole complementarity versus locality,''
  Phys.\ Rev.\  D {\bf 52}, 6997 (1995)
  [arXiv:hep-th/9506138].
}
\lref\GiNe{
  S.~B.~Giddings and W.~M.~Nelson,
  ``Quantum emission from two-dimensional black holes,''
  Phys.\ Rev.\  D {\bf 46}, 2486 (1992)
  [arXiv:hep-th/9204072].
}
\lref\WittenYR{
  E.~Witten,
  ``On string theory and black holes,''
Phys.\ Rev.\ D {\bf 44}, 314 (1991).
}
\lref\ADM{
  R.~L.~Arnowitt, S.~Deser and C.~W.~Misner,
 ``Canonical variables for general relativity,''
Phys.\ Rev.\  {\bf 117}, 1595 (1960).
}
\lref\Wal{
  R.~M.~Wald,
  ``On Particle Creation by Black Holes,''
Commun.\ Math.\ Phys.\  {\bf 45}, 9 (1975).
}
\lref\Unr{
  W.~G.~Unruh,
  ``Notes on black hole evaporation,''
Phys.\ Rev.\ D {\bf 14}, 870 (1976).
}
\lref\UQM{
  S.~B.~Giddings,
  ``Universal quantum mechanics,''
Phys.\ Rev.\ D {\bf 78}, 084004 (2008).
[arXiv:0711.0757 [quant-ph]].
}
\lref\BanksHST{
  T.~Banks,
  ``Holographic Space-Time: The Takeaway,''
[arXiv:1109.2435 [hep-th]].
}
\lref\CGHS{
  C.~G.~Callan, Jr., S.~B.~Giddings, J.~A.~Harvey and A.~Strominger,
  ``Evanescent black holes,''
Phys.\ Rev.\ D {\bf 45}, 1005 (1992).
[hep-th/9111056].
}
\lref\locbd{S.~B.~Giddings and M.~Lippert,
  ``Precursors, black holes, and a locality bound,''
  Phys.\ Rev.\  D {\bf 65}, 024006 (2002)
  [arXiv:hep-th/0103231];
  ``The information paradox and the locality bound,''
  Phys.\ Rev.\  D {\bf 69}, 124019 (2004)
  [arXiv:hep-th/0402073]\semi
   S.~B.~Giddings,
  ``Locality in quantum gravity and string theory,''
  Phys.\ Rev.\  {\bf D74}, 106006 (2006).
  [hep-th/0604072];
  ``(Non)perturbative gravity, nonlocality, and nice slices,''
  Phys.\ Rev.\  D {\bf 74}, 106009 (2006)
  [arXiv:hep-th/0606146]\semi
  S.~B.~Giddings, D.~Marolf,
  ``A Global picture of quantum de Sitter space,''
  Phys.\ Rev.\  {\bf D76}, 064023 (2007).
  [arXiv:0705.1178 [hep-th]].
}
\lref\GiddingsSJ{
  S.~B.~Giddings,
  ``Black hole information, unitarity, and nonlocality,''
Phys.\ Rev.\ D {\bf 74}, 106005 (2006).
[hep-th/0605196].
}
\lref\Jacrev{
  T.~Jacobson,
  ``Introduction to quantum fields in curved space-time and the Hawking effect,''
[gr-qc/0308048].
}
\lref\CavesZZ{
  C.~M.~Caves and B.~L.~Schumaker,
  ``New formalism for two-photon quantum optics. 1. Quadrature phases and squeezed states,''
Phys.\ Rev.\ A {\bf 31}, 3068 (1985).
}
\lref\GMH{
  S.~B.~Giddings, D.~Marolf and J.~B.~Hartle,
  ``Observables in effective gravity,''
Phys.\ Rev.\ D {\bf 74}, 064018 (2006).
[hep-th/0512200].
}
\lref\tHooftAA{
  G.~'t Hooft,
  ``A class of elementary particle models without any adjustable real parameters,''
Found.\ Phys.\  {\bf 41}, 1829 (2011).
[arXiv:1104.4543 [gr-qc]].
}
\lref\GiSl{
  S.~B.~Giddings and M.~S.~Sloth,
  ``Semiclassical relations and IR effects in de Sitter and slow-roll space-times,''
JCAP {\bf 1101}, 023 (2011).
[arXiv:1005.1056 [hep-th]];
``Cosmological observables, IR growth of fluctuations, and scale-dependent anisotropies,''
Phys.\ Rev.\ D {\bf 84}, 063528 (2011).
[arXiv:1104.0002 [hep-th]];
``Fluctuating geometries, q-observables, and infrared growth in inflationary spacetimes,''
[arXiv:1109.1000 [hep-th]].
}
\lref\nimatalk{N. Arkani-Hamed, talk at the KITP conference {\sl String phenomenology 2006.}}
\lref\ArkaniHamedKY{
  N.~Arkani-Hamed, S.~Dubovsky, A.~Nicolis, E.~Trincherini and G.~Villadoro,
  ``A Measure of de Sitter entropy and eternal inflation,''
JHEP {\bf 0705}, 055 (2007).
[arXiv:0704.1814 [hep-th]].
}
\lref\DVS{
  S.~Dubovsky, L.~Senatore and G.~Villadoro,
  ``Universality of the Volume Bound in Slow-Roll Eternal Inflation,''
[arXiv:1111.1725 [hep-th]].
}
\lref\Pageinfo{
  D.~N.~Page,
 ``Information in black hole radiation,''
Phys.\ Rev.\ Lett.\  {\bf 71}, 3743 (1993).
[hep-th/9306083].
}
\lref\PageDF{
  D.~N.~Page,
  ``Average entropy of a subsystem,''
Phys.\ Rev.\ Lett.\  {\bf 71}, 1291 (1993).
[gr-qc/9305007].
}
\lref\Sufast{
  L.~Susskind,
 ``Addendum to Fast Scramblers,''
[arXiv:1101.6048 [hep-th]].
}
\lref\STU{
  L.~Susskind, L.~Thorlacius and J.~Uglum,
  ``The Stretched horizon and black hole complementarity,''
Phys.\ Rev.\ D {\bf 48}, 3743 (1993).
[hep-th/9306069].
}
\lref\MaPl{
  S.~D.~Mathur and C.~J.~Plumberg,
  ``Correlations in Hawking radiation and the infall problem,''
JHEP {\bf 1109}, 093 (2011).
[arXiv:1101.4899 [hep-th]].
}
\lref\ModelsU{
  S.~B.~Giddings,
  ``Models for unitary black hole disintegration,''
[arXiv:1108.2015 [hep-th]].
}
\lref\HartGQM{
  J.~B.~Hartle,
  ``Space-time quantum mechanics and the quantum mechanics of space-time,''
[gr-qc/9304006].
}
\lref\AveryNB{
  S.~G.~Avery,
  ``Qubit Models of Black Hole Evaporation,''
[arXiv:1109.2911 [hep-th]].
}
\lref\GrLa{
  R.~Gregory and R.~Laflamme,
  ``The Instability of charged black strings and p-branes,''
Nucl.\ Phys.\ B {\bf 428}, 399 (1994).
[hep-th/9404071].
}
\lref\Sred{
  M.~Srednicki,
  ``Entropy and area,''
Phys.\ Rev.\ Lett.\  {\bf 71}, 666 (1993).
[hep-th/9303048].
}
\lref\BrPatra{
  S.~L.~Braunstein and M.~K.~Patra,
  ``Black hole evaporation rates without spacetime,''
Phys.\ Rev.\ Lett.\  {\bf 107}, 071302 (2011).
[arXiv:1102.2326 [quant-ph]].
}
\lref\GiddingsDR{
  S.~B.~Giddings,
  ``Is string theory a theory of quantum gravity?,''
[arXiv:1105.6359 [hep-th]]\semi
M.~Gary and S.~B.~Giddings,
  ``Constraints on a fine-grained AdS/CFT correspondence,''
[arXiv:1106.3553 [hep-th]].
}
\lref\vanr{
  M.~Van Raamsdonk,
  ``A patchwork description of dual spacetimes in AdS/CFT,''
Class.\ Quant.\ Grav.\  {\bf 28}, 065002 (2011); 
  ``Building up spacetime with quantum entanglement,''
Gen.\ Rel.\ Grav.\  {\bf 42}, 2323 (2010), [Int.\ J.\ Mod.\ Phys.\ D {\bf 19}, 2429 (2010)].
[arXiv:1005.3035 [hep-th]].
}
\lref\BanksTJ{
  T.~Banks,
  ``TASI Lectures on Holographic Space-Time, SUSY and Gravitational Effective Field Theory,''
[arXiv:1007.4001 [hep-th]].
}

\Title{
\vbox{\baselineskip12pt  
}}
{\vbox{\centerline{Black holes, quantum information, and unitary evolution}
}}
\centerline{{\ticp 
Steven B. Giddings\footnote{$^\ast$}{Email address: giddings@physics.ucsb.edu}  
} }
\centerline{\sl Department of Physics}
\centerline{\sl University of California}
\centerline{\sl Santa Barbara, CA 93106}
\vskip.10in
\centerline{\bf Abstract}
The unitary crisis for black holes indicates an apparent need to modify local quantum field theory.  This paper explores the idea that quantum mechanics and in particular unitarity are fundamental principles, but at the price of familiar locality.  Thus, one should seek to parameterize unitary evolution, extending the field theory description of black holes, such that their quantum information is transferred to the external state.  This discussion is set in a broader framework of unitary evolution acting on Hilbert spaces comprising subsystems.  Here, various constraints can be placed on the dynamics, based on quantum information-theoretic and other general physical considerations, and one can seek to describe dynamics with ``minimal" departure from field theory.  While usual spacetime locality may not be a precise concept in quantum gravity, approximate locality seems an important ingredient in physics. In such a Hilbert space approach an apparently ``coarser" form of localization can be described in terms of tensor decompositions  of the Hilbert space of the complete system.  This suggests a general framework in which to seek a consistent description of quantum gravity, and approximate emergence of spacetime.  Other possible aspects of such a framework -- in particular symmetries -- are briefly discussed.
 
\vskip.3in
\Date{}

\newsec{Introduction}

Lore holds that local quantum field theory (LQFT) is a unique reconciliation of the principles of quantum mechanics, 
Poincar\'e (more generally, diffeomorphism) symmetry, and locality.  Yet, in extreme regimes, these principles are in essential conflict.  In particular, we have strong indications that ultraplanckian collisions (or collapsing massive bodies) produce black holes.  Considering the fate of information inside such a black hole reveals the problem: each alternative -- the information escapes in Hawking radiation, the information is destroyed, or the information is left behind in a remnant -- contradicts one or more of these principles.  This conflict has been called the ``black hole information paradox," but here it will simply be called the unitarity crisis.\foot{For reviews, see \refs{\Astrorev\Pagerev\SGTrieste-\Mathurrev}.}

Such a basic clash indicates one or more principle requires modification, and relativistic LQFT is doomed.  Quantum mechanics and Lorentz symmetry seem very robust, and attempted modifications typically rapidly founder.  But, in quantum gravity, the concept of locality is remarkably hard to sharply formulate.\foot{For some further discussion of aspects of locality in gravity, see \refs{\Erice} and references therein.}

While difficult to formulate, locality is also hard to modify, and generic alterations to it in a framework including local quantum fields are expected to yield nonsense.  The reason is that, working in spacetime that is flat or nearly so, nonlocal transmission of information outside the light cone can be converted into transmission back in time.  This signaling into the past creates difficult paradoxes, which  seem to either indicate fundamental inconsistencies, or at least inconsistencies with the physics we observe.  Moreover, {\it local} QFT describes all experiments done and observations made to date, over an enormous span of scales, so locality apparently holds to a very precise approximation.

These observations indicate that if locality must be modified, there should be some more basic explanation of its validity, and of that of LQFT, as an excellent approximation to a deeper physics.

This paper will explore a possible resolution to these questions where quantum mechanics, suitably generalized,\foot{For comments on such generalization, see \refs{\UQM}.}  is taken  as fundamental, expanding ideas of \ModelsU.  This in particular indicates a Hilbert space description of ``physical degrees of freedom."  Then, at least in regimes where concepts like evolution and spacetime symmetries are good approximations, they are implemented by unitary transformations.  In particular, if locality and unitarity are in conflict, unitarity is assumed fundamental, and thus provides an important {\it constraint} on any possible dynamics.\foot{Note that {\it if} AdS/CFT were to provide a complete resolution of the problems of quantum gravity\GiddingsDR, these and related considerations of this paper may describe aspects of that resolution.}

In short, the viewpoint taken in this paper is that LQFT in spacetime is simply not the right fundamental picture, but that unitary ``evolution" on Hilbert space is.  One should then examine the consequences of this assumption.

Simply giving a Hilbert space and set of unitary transformations on it is insufficient.  Indeed, in practice, we essentially deal with finite-dimensional Hilbert spaces, and all such spaces of a given dimension are the same.  Extra structure is needed, starting with notions of localization, and continuing with evolution and symmetries.  This paper will outline first steps towards introducing such structure, in particular describing an analog to localization in factorization of Hilbert spaces into tensor factors.  It will also explore some basic features of evolution.  Of course, LQFT, where valid, yields unitary evolution on Fock space constructions of Hilbert space.  But, it has been argued that this physics cannot describe black hole evolution.  So, a natural place to explore the clash between locality and unitarity, and how it could be resolved in favor of unitarity, is in the black hole (BH) context. 

Specifically, the next section outlines some basic features of a possible Hilbert-space description of physics without fundamental spacetime, and in particular
 the possible relationship between locality and factorization of a Hilbert space into a network of tensor factors.  Next, section three reviews and refines the LQFT description of BH evolution, placing it in this more general Hilbert-space context, and providing the framework for its generalization.  Unitarity, then, indicates needed modification to this evolution.  Some basic constraints on, and models for, unitary evolution laws are described in section four.  In particular, this evolution must transfer quantum information from BH states into external states, and basic aspects  of quantum information theory together with the desire to provide a close match to LQFT evolution (or, if some views are believed, not) suggest various important constraints.  Implications of these constraints are examined in models for evolution, and an important problem for the future is to further refine such models and other constraints on them.  Section five continues general discussion of how to describe basic features of physics in such a Hilbert space framework, and in particular comments on the implementation of symmetries, global and local, and on other guides to a fundamental mechanics within the framework of such a Hilbert space network.

\newsec{Hilbert-space networks and gravitational mechanics}

This paper will explore the viewpoint that essential features of quantum mechanics are fundamental -- minimally the basic structure of Hilbert space.  Symmetries, and particularly Poincar\'e symmetry when describing an asymptotically Minkowski system, are implemented by unitary transformations; for some further discussion see section five.  This leaves the question of locality.

In local quantum field theory, locality is encoded in the statement that gauge-invariant local operators commute outside the lightcone:
\eqn\commut{[\calo(x),\calo(y)]=0\quad,\quad (x-y)^2>0\ .}
This statement can be extended to localized operators; consider, {\it e.g.}, Wilson loops with support in spacelike-separated regions.  In the geometrical description of gravity, there are no gauge-invariant local operators, since a diffeomorphism acts as 
\eqn\diffact{\delta_\xi \calo(x) = \xi^\mu \partial_\mu \calo(x)\ .}
Yet, we expect locality to emerge to an excellent approximation for systems described in weakly-gravitating semiclassical spacetime.

When considered in Fock-space terms, \commut\ is the statement that operators in spacelike separated regions create independent tensor factors of the Hilbert space of states: action of an operator here does not influence simultaneous action of an operator in the Andromeda galaxy.  While quantum gravity is not expected to have precisely this kind of spacetime locality -- particularly if it is to describe unitary evolution of black holes -- a plausible hypothesis is that it does have a possibly coarser form of locality implemented through factorization of its Hilbert space into tensor factors, and through constraints on the action of the evolution on such factors.

 \Ifig{\Fig\Tensfact}{A schematic of factorization of a large Hilbert space into tensor factors.  Intersections correspond to common factors of the overlapping Hilbert spaces.}{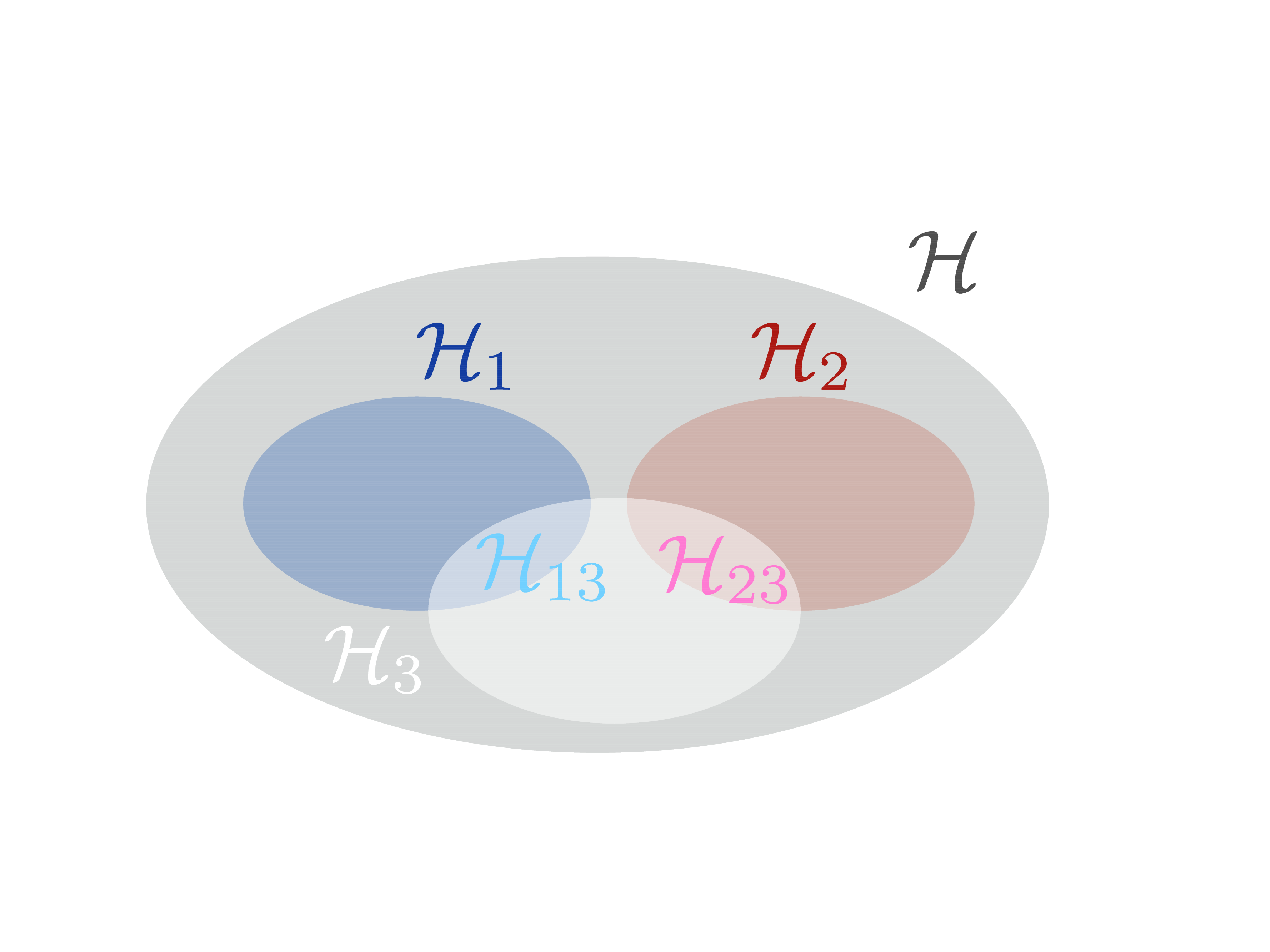}{4.5}

Specifically, consider \Tensfact.  We can think of independent tensor factors $\calh_1$ and $\calh_2$ of the total Hilbert space $\calh$ as corresponding to different regions of space.  One can also have two different factors with a common factor, as with $\calh_1$ and $\calh_3$, in which case the common factor $\calh_{13}$ corresponds to the intersection of the regions.  It should be stressed that in this viewpoint, the structure of the net of overlapping/nested Hilbert spaces is taken as basic, and no additional spacetime manifold structure is assumed.  Spacetime emerges to the extent that it gives a good approximate description of the Hilbert tensor network and its evolution.  A question for the future is to further explore different such network structures, and their relation to approximate spacetimes.  Note also, that for describing physics in a finite region, and at finite total energy, one might in practice consider finite dimensional Hilbert spaces.\foot{However, implementation of Lorentz symmetry does require an infinite-dimensional factor associated with arbitrarily large total momentum of a system.}  

Such a viewpoint is not completely novel.  In fact, in the algebraic approach to LQFT,\foot{For a review, see \refs{\Haag}.} the basic framework is that of a net of algebras of local observables, associated with causal diamonds in spacetime.  These observables can be thought of as acting within tensor factors.  For the purposes of the present discussion it is convenient to focus on the latter structure, although factorization of the operator algebra and the Hilbert space are clearly related.

There are, however, important differences in the present approach.
 In algebraic quantum field theory, background spacetime structure is {\it assumed}.  In the approach being advocated here, the basic structure is that of the tensor network of Hilbert space factors, and associated unitary dynamics.  This is assumed to be a valid way to describe the fundamental physics {\it also incorporating gravity.}  If such a structure indeed captures the correct physics, then it provides a way that dynamical geometry could emerge, in an approximation, instead of being introduced at the start. 
The fact that there is a smallest non-trivial Hilbert space, of two dimensions, strongly suggest a coarser structure than that of a manifold, and one that does not necessarily incorporate the notion of 	``points" in spacetime.  This viewpoint is also reinforced if one restricts attention to finite-dimensional Hilbert spaces.

A structure based on causal diamonds, together with a description of world lines and associated local observers, the notion of ``holographic screens," and an essential reliance on implementation of supersymmetry, also serves as a cornerstone of Banks' proposed ``holographic spacetime."\foot{For reviews and further references see \refs{\BanksHST,\BanksTJ}.}  This has common features with the present approach, specifically in describing physics in terms of a network of Hilbert spaces, but also differs from the present proposal in a number of respects, particularly the input assumptions of these various elements of specific extra spacetime and physical structure.  Also, that approach associates small Hilbert spaces with ``pixels" on a holographic screen, rather than regions of spacetime, and thus is intrinsically ``holographic." (Related structures have been considered in \refs{\vanr}.)

One other consequence of the present approach is that, since evolution is implemented by unitary transformations on Hilbert space, information transfer is constrained by the basic principles of quantum information theory:  all information is quantum information.  This offers important constraints on the framework, which will emerge in describing physical systems such as black holes.

\newsec{Local quantum field theory: Hawking evolution and tensor factorizations}

In order to sharpen these ideas,  this section will give a detailed treatment of Hawking evolution.  While this  is ultimately nonunitary\Hawkunc, its formulation will both illustrate implementation of a more general Hilbert-space description, and set the stage for finding modifications to the evolution that respect unitarity, within such a description.  We begin by updating the derivation of Hawking radiation, based on a treatment of LQFT evolution on a time slicing of a black hole spacetime.

\subsec{Geometry and coordinates}

Consider, in particular, an asymptotically-flat spacetime in which we imagine a black hole forms through a collapse or collision from a pure initial state.  For simplicity we will take the geometry to be spherically symmetric, although there is also much interest in high-energy collisions with less symmetry\refs{\Erice}.  The general spherically-symmetric D-dimensional black hole metric is
\eqn\schmet{ds^2 = -f(r) dt^2 + {dr^2\over f(r)} + r^2 d\Omega^2\ .}
For the Schwarzschild solution,
\eqn\schrad{f(r)= 1-\left({R\over r}\right)^{D-3}\ ,}
with Schwarzschild radius $R$.\foot{In $D$ dimensions, $R$ is given in terms of the mass $M$ and Newton's constant $G_D$ as $R^{D-3}=\{8\Gamma[(D-1)/2)]\}/[\pi^{(D-3)/2}(D-2)] G_D M$.} 

It is useful to introduce new coordinates to give a simultaneous description of the black hole interior and exterior.  First, one defines tortoise coordinate $r_*$, making the $r,t$ plane manifestly conformally flat:
\eqn\tortmet{ds^2 = f(r) (-dt^2 + dr^{*2}) + r^2(r_*) d\Omega^2\ ,}
with the definition
\eqn\tortdef{r_* = \int {dr\over f(r)}\ .}
The coordinates $r$ and $r_*$ match asymptotically, but $r_*=-\infty$ at the horizon $r=R$.  It is also useful to define null coordinates
\eqn\xcoords{x^{\pm} = t\pm r_*\ .}
Eq.~\tortmet\ describes the black hole exterior, and the interior is found by using Kruskal coordinates, {\it e.g} in the null form
\eqn\Kruskdef{X^{\pm}  = \pm R e^{\pm f'(R)x^\pm/2}\ }
Then, the future horizon is $X^-=0$, and the resulting expression for the metric continues to the singularity at $r=0$.  The singularity is also given by 
\eqn\singcurv{X^+X^-=R_0^2\ ,}
 where $R_0$ is a dimension-dependent constant times $R$, and can be determined from formulas in \refs{\GrLa}.
A particularly simple example of such a black hole geometry is that of the two-dimensional black hole\refs{\WittenYR,\CGHS},
\eqn\tdsoln{ds^2_{2dBH}= -{dX^+dX^-\over 1-\lambda^2 X^+X^-}\ ,}
where $\lambda$ is a constant of dimension inverse length.

\Ifig{\Fig\Kdiag}{A Kruskal diagram  for a black hole.  Also shown is matter infalling in the far past; if the black hole formed from this matter, the vacuum geometry below it, including $X^+\leq0$, is not present.}{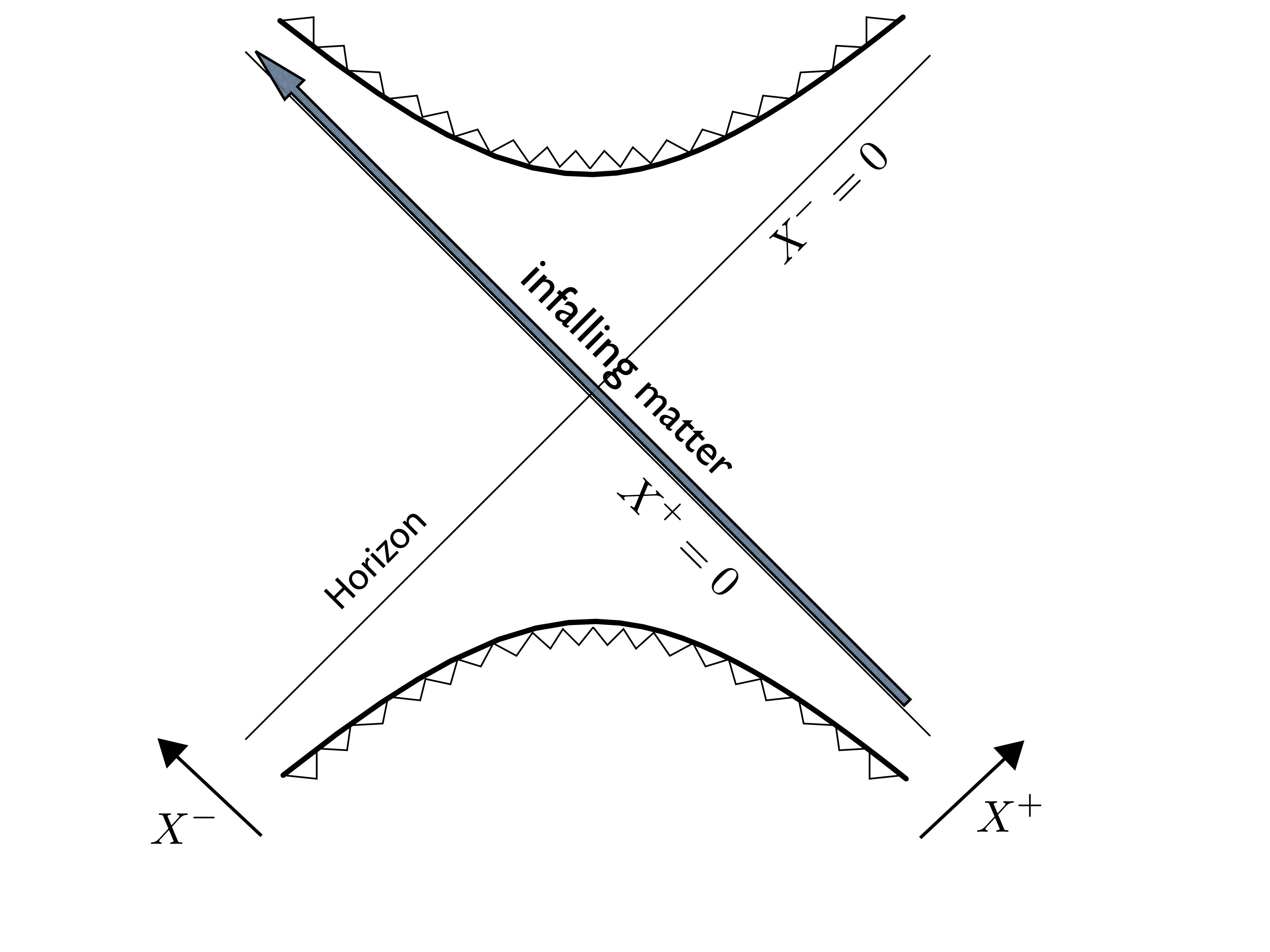}{4.5}

\Ifig{\Fig\EddFink}{A black hole pictured in Eddington-Finkelstein coordinates, together with a family of nice slices, and schematic representation of production/evolution of paired Hawking excitations.}{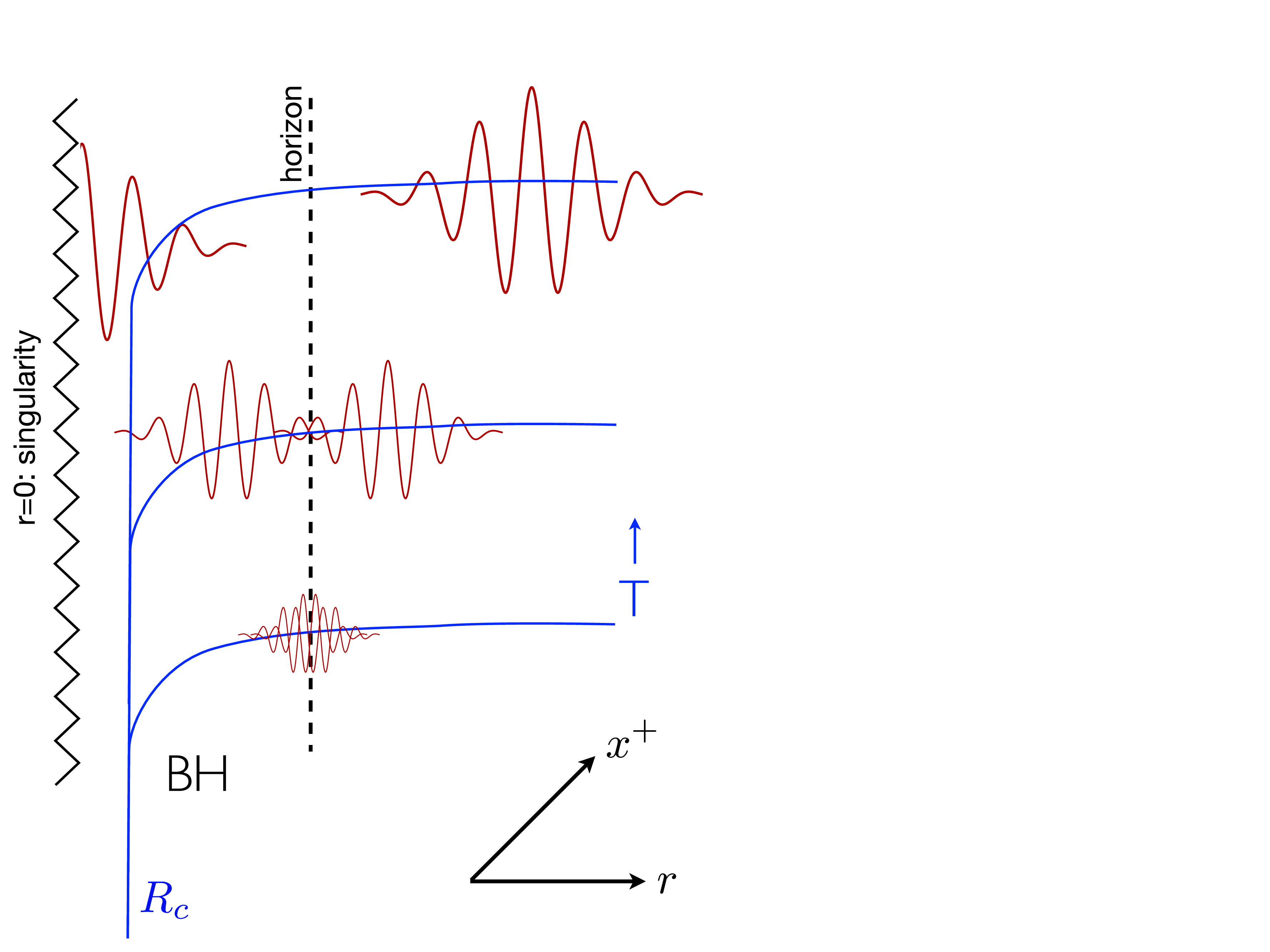}{3.5}

For a black hole formed from infalling matter, the lower portion of the Kruskal manifold, and  in particular $X^+\leq0$, is absent.  For infalling matter in the far past, the Kruskal diagram is pictured as in \Kdiag.  Another way to visualize the geometry of collapse and latter black hole evolution is in Eddington-Finkelstein coordinates, labeling points by $(x^+, r, \Omega)$, and an Eddington-Finkelstein picture of the geometry is \EddFink.

\subsec{LQFT evolution}

Dynamics in a spacetime is conveniently treated by choosing a time slicing, and this in particular helps us to give a representation of the Hilbert space of states, and decompositions into tensor factors.   The metric in such a slicing takes the ADM form\refs{\ADM},
\eqn\admdec{ds^2=-N^2 dT^2 + g_{ij} (dx^i + N^i dT)(dx^j + N^j dT)}
where $g_{ij}$ is the spatial metric within a slice of constant time $T$, and $N,N^i$ are the lapse and shift.  A simple and readily-generalizable example of quantization on such slices is that of the massless scalar field, with lagrangian
\eqn\slagr{{\cal L} = - \hf g^{\mu\nu} \partial_\mu \phi \partial_\nu \phi}
The canonically-conjugate momentum to $\phi$ is 
\eqn\canonmom{\pi = {\partial_T \phi - N^i \partial_i\phi\over N}= n^\mu\partial_\mu\phi\ ,}
where $n^\mu$ is the normal to the slice,
and satisfies
\eqn\CCR{[\pi(x,t),\phi(y,t)] = -i{\delta^{D-1}(x-y)\over \sqrt{{}^{D-1}g}}\quad \leftrightarrow \quad \pi= -i {\delta \over \delta\phi}\ .}

The hamiltonian is given by
\eqn\hgen{N{\cal H} = \hf N(\pi^2 + g^{ij}\partial_i \phi \partial_j \phi )+ N^i \pi\partial_i\phi \ , }
on a fixed background the unitary evolution operator of LQFT is
\eqn\unitop{U_{LQFT}=\exp\left\{ -i \int dt d^{D-1} x\sqrt{{}^{D-1}g}  N{\cal H}\right\}\ .}
These expressions also extend to higher-spin theories, and interacting theories, through generalization of \hgen\ and of  the canonical structure.

\subsec{Example slicings: natural, nice, and Schwarzschild}

It is helpful to have explicit descriptions of slicings for the static black hole metric \schmet. Such a slicing is given by solutions to the equation
\eqn\sldef{X^+(X^-+e^{-f'(R)T} X^+) = R_c^2\ ,}
for a fixed $R_c>0$.  Different slicings are given by different $R_c$.  The case $R_c=0$ gives the Schwarzschild slicing, where the slices never cross the horizon.  For $R_0>R_c>0$ (with $R_0$ as in \singcurv), we get an example of a ``nice slicing\refs{\LPSTU}," which agrees with Schwarzschild time slices at infinity, but in which the slices smoothly enter the horizon, but not the region $r<R_c$, and in particular avoid the singularity.  In the case $R_c>R_0$, the slices intersect the singularity; these were referred to as ``natural slices" in \refs{\NLvC} since they could describe time slices defined by a family of infalling observers.

For slices like these that respect the radial symmetry of Schwarzschild \schmet, the ADM form of the metric \admdec\ simplifies to
\eqn\admrad{ds^2 = -N^2 dT^2 + g_{xx} (dx+N^x dT)(dx+N^x dT) + r^2(T,x)d\Omega^2\ }
where $x$ is a coordinate parameterizing the radial direction along the slice.  One such choice is $x=X^-$; for example, given this, slices described by \sldef, and the Schwarzschild metric, the coefficients in \admrad\ may be computed explicitly.

Particular focus here will be on the nice slicing, as it cleanly exhibits the tension between LQFT and unitarity of black hole evolution.  One can readily check\QBHB\ that in such a slicing ({\it e.g.} as just described), the lapse $N$ asymptotes to zero at the inner boundary $r=R_c$.  Thus LQFT evolution described by $U_{LQFT}$ of \unitop\ corresponds to {\it freezing} of the state at $r=R_c$, as is familiar from the case $R_c=0$.  This furnishes a convenient description of the state of matter that has fallen into a black hole -- the state is frozen into a ``record" laid out along the nice slice.   As we will note later, this also provides a way to describe the $\sim \exp\{S_{BH}\}$ states expected for a black hole,
where 
\eqn\sbhdef{S_{BH}=  {R^{D-2}\Omega_{D-2}\over 4 G_D}}
is the Bekenstein-Hawking entropy.  ($\Omega_{D-2}$ is the volume of the unit $D-2$ -sphere.)

\subsec{Position-momentum localization and tensor product description}

The previous two sections have set up a Hilbert-space description of LQFT evolution, based on a Fock space construction on a curved background, using a definite slicing.  This section will provide further details, and indicate how one finds a tensor-product structure corresponding to different regions.

Specifically, in describing such a tensor product structure,\foot{This kind of tensor decomposition has been applied in various related contexts; see {\it e.g.} \refs{\Sred} or more recently \refs{\BrPatra}.}  it is in general useful to separate modes according to their momenta transverse to and parallel to the boundary separating the regions.  This section gives an example with boundary which is  the horizon of the spherically-symmetric black hole \schmet.  Here, it is convenient to expand the general solution of \slagr\ in terms of spherical harmonics,
\eqn\spherexp{\phi = \sum_{l\vecm}{\phi_{l,\vecm}(r,t)\over r^{D/2-1}} Y_{l\vecm}(\Omega)\ .  }

To proceed, we describe modes localized in both momentum and position (here radial).  In general there are many ways to choose a basis of such modes.  For simplicity, these may be chosen to be orthonormal in the inner product
\eqn\innerprod{(\phi_1,\phi_2)= i\int_\Sigma d^{D-1}x\sqrt{ {}^{D-1}g} n^\mu \phi_1^*{\overleftrightarrow {\partial_\mu}} \phi_2\ ;}
on a constant-$t$ slice this simplifies to
\eqn\innerprodt{(\phi_1,\phi_2)=i\sum_{l\vecm} \int dr_*\phi_{1;l,\vecm}^*{\overleftrightarrow {\partial_t}}\phi_{2;l,\vecm}   \ .}

For example, on a time-slice described by the coordinatization \admrad, we can, for each $l$, choose a basis of localized functions $u_{il}(x)$, so that the field is expanded as
\eqn\tdexpan{\phi(x,T,\Omega)=\sum_{il\vecm}\left(a_{il\vecm} u_{il} {Y_{l\vecm}(\Omega)\over r^{D/2-1}} + {\rm h.c.}\right)\ .}
One way to do this is to superpose basis functions 
\eqn\planebase{\exp\{\pm i k x\}\ ,}
 and one simple basis of localized modes is given by the windowed Fourier transform,
\eqn\uki{u_{ja} = {1\over \sqrt\epsilon} \int_{j\epsilon}^{(j+1)\epsilon} dk  e^{ik(x-2\pi a/\epsilon)}\ ,\ {\tilde u}_{ja} =  {1\over \sqrt\epsilon} \int_{j\epsilon}^{(j+1)\epsilon} dk  e^{-ik(x-2\pi a/\epsilon)}}
with $\epsilon$ an arbitrarily-chosen resolution parameter.\foot{Such modes have been used in studies of Hawking radiation in \refs{\Hawkrad,\GiNe}; see {\it e.g.} the latter for further properties. }  Such localized modes may then be extended to positive-frequency solutions, in a chosen convention.  While the solutions extending a general basis  $u_{il}(x)$ are not necessarily a priori orthogonal under \innerprod, one expects to be able to form an orthogonal basis {\it e.g.} via a Graham-Schmidt procedure.  Moreover, in the large-momentum limit where the metric \admrad\ appears approximately flat, the modes corresponding to \uki\ are orthogonal.  
Next,  if the modes are divided up into those (approximately) localized inside or outside the black hole, the corresponding decomposition of the Fock space specifies a Hilbert-space decomposition on the given time slice,
\eqn\hprod{\calh(T) = \hbh(T)\otimes\hext(T)\ .}

This procedure becomes particularly clear in both the near-horizon and far-field limits, where the dynamics reduces to that of a two-dimensional model.  In the massless two-dimensional model, for example with $x=X^-$, the basis \planebase\ extends to 
 positive Kruskal-frequency solutions with $\omega=k$ and matching signs in the exponents:
\eqn\kruskbasis{e^{-i\omega X^-}\ ,\ e^{-i\omega X^+(X^-,T)}\ .}
The modes $u_{ja}$, ${\tilde u}_{ja}$ of \uki\ have approximately definite momenta $\pm k\simeq\pm j\epsilon$ and positions $x=X^-\simeq 2\pi a/\epsilon$.  The Hilbert space on a given time slice may be decomposed as in \hprod\ according to whether $a>0$ or $a<0$.

A sharper distinction between subsystems, important in deriving the Hawking radiation, is to construct the outward-moving modes using instead the coordinate $x^-$.  There is an analogous coordinate describing the region inside the horizon, given by (compare \Kruskdef)
\eqn\xhmdef{{\hat x}^- = {2\over f'(R)} \ln (X^-/R)\ .}
Outgoing positive Schwarzschild frequency analogs of \kruskbasis\ are 
\eqn\schwbasis{v_\omega = e^{-i\omega x^-}\ ,\ {\hat v}_\omega =e^{-i\omega{\hat x}^-}\ ,}
supported only in the regions $X^-<0$ and $X^->0$, respectively.  Then, these can be superposed to form wavepackets that are further localized, as in \uki\ (though now including time dependence):
\eqn\vki{v_{ja} =  {1\over \sqrt\epsilon} \int_{j\epsilon}^{(j+1)\epsilon} d\omega  e^{-i\omega(x^--2\pi a/\epsilon)}\ ,\ {\hat v}_{ja} =  {1\over \sqrt\epsilon} \int_{j\epsilon}^{(j+1)\epsilon} d\omega  e^{-i\omega({\hat x}^--2\pi a/\epsilon)}\ ,}
and these, plus ingoing modes, provide an alternate decomposition \hprod\ of the total Hilbert space.  Again these give just one example of a general choice of wavepacket basis $v_{il}$, ${\hat v}_{il}$, ${\tilde v}_{il}$.

{\it Arbitrary} Fock space states in a product like \hprod\ are not expected to correspond in any simple way to physical states. For example, excitation of two ultraplanckian modes in a small region produces a strong gravitational backreaction.  A proposed quantification of such limitations is the locality bound\refs{\locbd,\GiddingsSJ}.  
And, a black hole that formed in the far past has been argued to be in a state well-approximated as the Unruh vacuum\refs{\Unr}. This is found by decomposing the modes into those that are in the far past outgoing or ingoing. Then, this state is vacuum with respect to outgoing modes that are positive frequency in Kruskal time, and ingoing modes that are positive frequency in Schwarzschild time. 
In the two-dimensional (near-horizon) approximation, the Unruh vacuum can thus be represented as 
\eqn\unrstate{|0\rangle_U = |0\rangle_{x^+} |0\rangle_{X^-}\ .}
Then, acting with creation operators corresponding to positive frequency (in $x^+$)  ingoing modes produces more general states with infalling matter. The Unruh state, or more general state with infalling matter, can be decomposed in a tensor product \hprod; the space of such states forms a subspace of the product Hilbert space.

These states evolve via the unitary operator \unitop.   Defining this requires a normal-ordering prescription, with respect to the positive-frequency modes, and thus in particular with respect to the Kruskal modes for outgoing states at the horizon.  The evolution may be generalized to interacting fields, though we focus on the free case for simplicity.

For the purposes of describing observations of asymptotic observers, one works in a basis like \vki, constructed using the coordinates $x^-$, ${\hat x}^-$.  These modes are related to the Kruskal modes by a Bogolubov transformation, and in this basis the Unruh vacuum has an infinite number of particles.

The latter  feature is readily understood using a trick due to Wald\refs{\Wal}.\foot{For further details on the relation between the states, see \refs{\GiNe,\SGTrieste,\Jacrev}.}  Positive Kruskal-frequency modes are analytic in the lower-half complex $X^-$ plane.  So, the following functions are positive-frequency
\eqn\cplxmod{\eqalign{u^+_\omega=(X^-)^{-2i\omega/f'(R)}&\propto {\hat v}_\omega + e^{-\beta\omega/2} v^*_\omega\cr
u^-_\omega=(-X^-)^{2i\omega/f'(R)} &\propto v_\omega+ e^{-\beta\omega/2} {\hat v}^*_\omega\ ,}}
with
\eqn\BHtemp{\beta={4\pi\over f'(R)} = {4\pi R\over D-3}\ ,}
and their corresponding annihilation operators must annihilate the Unruh state. If $b_\omega$, ${\hat b}_\omega$ are the annihilation operators associated with the modes $v_\omega$ and ${\hat v}_\omega$, this is the condition that the state is annihilated by
\eqn\statecond{\eqalign{&b_\omega-e^{-\beta\omega/2} {\hat b}_\omega^\dagger\cr& {\hat b}_\omega-e^{-\beta\omega/2} { b}_\omega^\dagger\ .}}
This then determines the state in the $v_\omega,{\hat v}_\omega$ basis.  Introducing the occupation-number basis for the number operators $b_\omega^\dagger b_\omega$ and ${\hat b}_\omega^\dagger {\hat b}_\omega$, 
this state may be written, somewhat formally, as
\eqn\stateexp{|0\rangle_{X^-} = {1\over\sqrt{Z}} \sum_{\{n_\omega\}} e^{-{\beta\over 2} H} |\widehat{\{n_\omega\}}\rangle|\{n_\omega\}\rangle = \prod_\omega S(\omega)|\hat 0\rangle|0\rangle\ .}
Here $H$ is the Schwarzschild hamiltonian,
\eqn\schwH{H=\int_0^\infty d\omega \omega n_\omega}
and
$Z$ is a normalization factor.  $S(\omega)$ is a unitary squeeze operator\CavesZZ\ for modes at $\omega$, 
\eqn\sqedef{S(\omega) = \exp\left\{ {z(\omega)} \left( {\hat b}_\omega^\dagger b_\omega^\dagger -  {\hat b}_\omega b_\omega\right)\right\}\ ,}
with
\eqn\zdef{\tanh z(\omega) = e^{-\beta \omega/2}\ .}

Eq.~\stateexp\ (together with the decomposition of the ingoing state $|0\rangle_{x^+}$, or excitation thereof, described previously) does express the state of a black hole in the product form \hprod.  It is formal, however, due to inclusion of an unphysical infinity of modes.  To regulate it, we first go to a basis of position-momentum localized modes, like the example \vki\ above.

Specifically, working on a fixed time slice labelled by $T$, superpose solutions  to give an orthonormal basis of position-momentum localized wavepackets.  In particular, the near-horizon, instantaneously-outgoing modes can be approximated as superpositions of \cplxmod, {\it e.g.} extending \vki.  In $D>2$, one can also include angular momentum.  Thus, the obvious generalization of \stateexp\ is
\eqn\statepack{|0\rangle_{X^-} = {1\over\sqrt{Z}} \sum_{\{n_{jal}\}} e^{-{\beta\over 2} H} |\widehat{\{n_{jal}\}}\rangle|\{n_{jal}\}\rangle \ ,}
where $H$ is re-expressed in the basis $v_{jal}$.  Or, one may extend the expression to a more general basis $v_{il}$. Indeed, note that if we use the modes \vki\ and choose $\epsilon\ll 1/R$ the state is annihilated by the analogous combinations to \statecond, with $\omega$ replaced by $\omega_j$, and the state can also be written
\eqn\sqopvers{|0\rangle_{X^-} =\prod_{jal} S_{jal} |\hat 0\rangle|0\rangle}
with
\eqn\dissque{ S_{jal} = \exp\left\{{z(\omega_j)} \left( {\hat b}_{jal}^\dagger b_{jal}^\dagger -  {\hat b}_{jal} b_{jal}\right)\right\}\ .}
In a more general basis, the expression in this exponential is non-diagonal, but of a similar form.

Contributions of modes with $\omega_j\gg 1/R$ are exponentially suppressed by the thermal factors, providing an effective cutoff.  
However, there is still an infinity in \statepack, \sqopvers\ from the range of $a$.  To regulate this, note that for a given $T$, modes with sufficiently large $a$ have wavelength $\ll R$, and are localized a comparably small separation from the horizon.  Indeed, while these modes can have high, even ultraplanckian, energies as seen by an infalling observer, the statement that they are in a {\it paired} state corresponding to $|0\rangle_{X^-}$  means that potentially large interactions between infalling matter and the individual modes cancel between members of a pair\refs{\GiddingsSJ}. So, we rewrite \statepack\ or \sqopvers\ by restricting the range of $a<A(T)$, where $A(T) =  \epsilon(T+kR)/(2\pi)$, and $k(L)$ is chosen in order that modes whose slice distance to the horizon is less than a cutoff value $L$ are not included. 
The contribution of the latter modes can be rewritten as $|0\rangle_{A(T)}$, representing the fact that for practical purposes the modes are seen as vacuum; in particular
\eqn\sqopversc{|0\rangle_{X^-} =\prod_{jl}\prod_a^{A(T)} \left(S |\hat 0\rangle|0\rangle\right)_{jal} |0\rangle_{A(T)}}
and analogously for \statepack\ (here the vacuum is also decomposed by mode).

At time $T$ eq.~\sqopversc\ provides a decomposition of the ``outgoing" near-horizon states of the product form \hprod; the part of the state giving ingoing matter may be decomposed as described below \kruskbasis.  The factor $|0\rangle_{A(T)}$ is one-dimensional, and largely trivial; it may be for example associated with $\calh_{BH}$.

Evolution to a later time $T'$ is given by the unitary operator $U(T',T)$, defined in \unitop, which can also be generalized to the case of interacting fields.  This operator describes motion of outgoing modes outside the horizon away from the black hole, and other modes towards the black hole center.  At the same time, since near-horizon modes separate from the horizon, a constant ``physical" cutoff corresponds to an evolving $A(T)$:
\eqn\vacevolve{|0\rangle_{A(T)} = \prod_{jl}\prod_{a=A(T)}^{A(T')} \left(S |\hat 0\rangle|0\rangle\right)_{jal}|0\rangle_{A(T')}\ .}
As a result of this, the factor Hilbert spaces in \hprod\ change with time.  However, they do so such that $U$ linearly maps between physical states in one-to-one fashion, preserving the inner product, and thus can be described as unitary in this generalized sense.

In $D>2$ or for non minimally-coupled matter, another source of rearrangement between the factors during time evolution is reflection; {\it e.g.} an initially outgoing near-horizon mode can, through evolution with $U(T',T)$, become a superposition of an outgoing mode and an ingoing mode that falls into the horizon.  In fact, asymptotic outgoing modes with $l\gg \omega R$ are strongly suppressed precisely through such factors.  At the later time $T'$, we then may choose a mode basis that describes the localization of the particles at that time.  

\subsec{Hawking radiation and breakdown of unitarity}

To describe the state outside the black hole, we trace over $\calh_{BH}(T)$ in our expressions for $|0\rangle_U$.  For example, neglecting reflection,\foot{More generally, in free evolution we expect $v_{jal}(T)\rightarrow R_{jal}\, {\tilde u}_{BH,{ja'l}}(T') + T_{jal}\, v_{ext,{jal}}(T')$, with computable reflection and transmission coefficients.  The transmission coefficients, or ``gray-body factors," then enter the following expression for an asymptotic density matrix.} and focusing on the expression \statepack\ for outgoing modes, this becomes
\eqn\densmat{\rho(T) = {1\over Z} \sum_{\{n_{jal}\}, a<A(T)}  e^{-\beta H} |\{n_{jal}\}\rangle\langle\{n_{jal}\}|\ ,}
a thermal density matrix.  (Additionally, modes of infalling matter can contribute.)  As more modes cross the cutoff at $a=A(T)$, the rank of the density matrix increases. Missing information in the density matrix is characterized by the von Neumann entropy,
\eqn\vN{S(T)= -{\rm Tr}[\rho(T)\log \rho(T)]\ .}
If one neglects backreaction of the Hawking radiation on the geometry, this entropy grows to size $S_{BH}$ on the black hole evaporation timescale, 
\eqn\Tbh{T_{BH}\sim R S_{BH}\ .}
The large missing information quantifies the violation of unitarity first proposed by Hawking\refs{\Hawkrad,\Hawkunc}.

\subsec{Modifications from dynamical gravity}

The  state of the Hawking radiation, and specifically the Unruh vacuum, has so far been described for a black hole of constant mass.  This is an extremely good approximation for Hawking radiation from a large black hole, since of order one Hawking quantum of energy $1/R$ is emitted each time $T\sim R$, and so the black hole mass only varies by an order one fraction on the timescale $T_{BH}$.
On shorter timescales, the mass is essentially constant.  Thus, the localized features of the state, such as the division into modes in a given small interval of space or time, and the state of those modes, appear largely insensitive to this variation.  In particular, the existence of decompositions into tensor factors, such as \hprod, appears valid.  The effect of the changing geometry does alter the space of states and its evolution, for example as seen through appearance of $(N,N^i,g_{ij})$ in \unitop, or its interacting generalizations.  
And, over scales $\sim T_{BH}$, one expects significant distortion of the state, due to the change in the geometry.

For observations on small portions of a spatial slice through the geometry, or even over large portions in the asymptotic region, one can imagine promoting slice-dependent (thus gauge-dependent) statements into gauge-invariant statements by introduction of extra structure\refs{\GMH} corresponding to a ``reference background,"  or  set of ``local observers."  However, as noted in \QBHB, giving such a gauge-invariant characterization of the entire nice-slice state, over times $\sim T_{BH}$, is problematic.  The reason is that a reference background with sufficient resolving power over an interval of size $T$ on the slice must itself carry an energy $E\roughly> T/R^2$ (corresponding to a minimum of one quantum of wavelength $R$ per time $R$), and so by time $\tbh$ causes an order one perturbation.  

One can try to ignore this, and attempt an ADM quantization on the nice slices.  In particular, this leads to a generalization of \unitop\ including interacting perturbations corresponding to gravitational fluctuations\QBHB. A first problem in this is specifying a gauge condition on the slices that yields well-defined evolution, once perturbations away from the metric \schmet\ become important.  In-depth treatment of such quantization is left for further work, but general features appear.  First, as was argued in \QBHB, it appears that such a perturbative quantization becomes problematic, by times $\sim\tbh$, due to gravitational coupling between the fluctuations.  If, as argued there, this indeed represents a breakdown of any perturbative derivation of the late-time nice slice state, there is no sharp calculation of the missing information.  Indeed, a similar but simpler context to explore perturbative ADM quantization of a curved geometry is in inflationary cosmology; here one has more symmetry.  Interestingly, one also finds an apparently related breakdown of perturbative quantization there\refs{\QBHB,\GiSl}\foot{For a different, but likely connected story, see \refs{\nimatalk\ArkaniHamedKY-\DVS}.}

In short, a proposed\QBHB\ resolution to the ``information paradox" is that the nice-slice state does not accurately describe the quantum state of a black hole at long times.  In particular, 
it appears that we do not have a gauge-invariant and perturbatively-sound sharp calculation of the missing information, past the time $\tbh$, and if there is no such calculation, there is no sharp paradox.

Such a resolution does not yet offer the full story, however, as one needs a more complete description of the correct, presumably unitary, evolution.  Apparently this must involve both a modification of the LQFT description of the nice slice state, and of its evolution via \unitop.  We seek a set of principles governing such evolution.

In investigating such possible principles, this paper will assume some basic structure, even in the presence of the quantum generalization of a ``fluctuating metric."  Particularly, description of gravitating systems by Hilbert spaces will be assumed, to implement quantum mechanics, as will the existence of decompositions of these into tensor factors, which then correspond to different regions in the limit where semiclassical geometry is recovered.  One approach to motivating the latter is to note that it can be made asymptotically on the states, and then the states can be evolved {\it e.g.} adiabatically into the regions in question.  An ultimate test of these assumptions is whether they are indeed consistent with formulation of a dynamics which consistently describes gravitating systems.  We will begin to explore this by investigating possible unitary modifications to LQFT evolution, in the context of an evaporating black hole.

\newsec{Unitary evolution}

\subsec{Basic considerations and expectations}

\Ifig{\Fig\entropfig}{A sketch of the growth of the von Neumann entropy of the external state resulting from Hawking evolution, as compared to its ultimate decline (here pictured as beginning near the half life of the black hole) necessary for unitary evolution.}{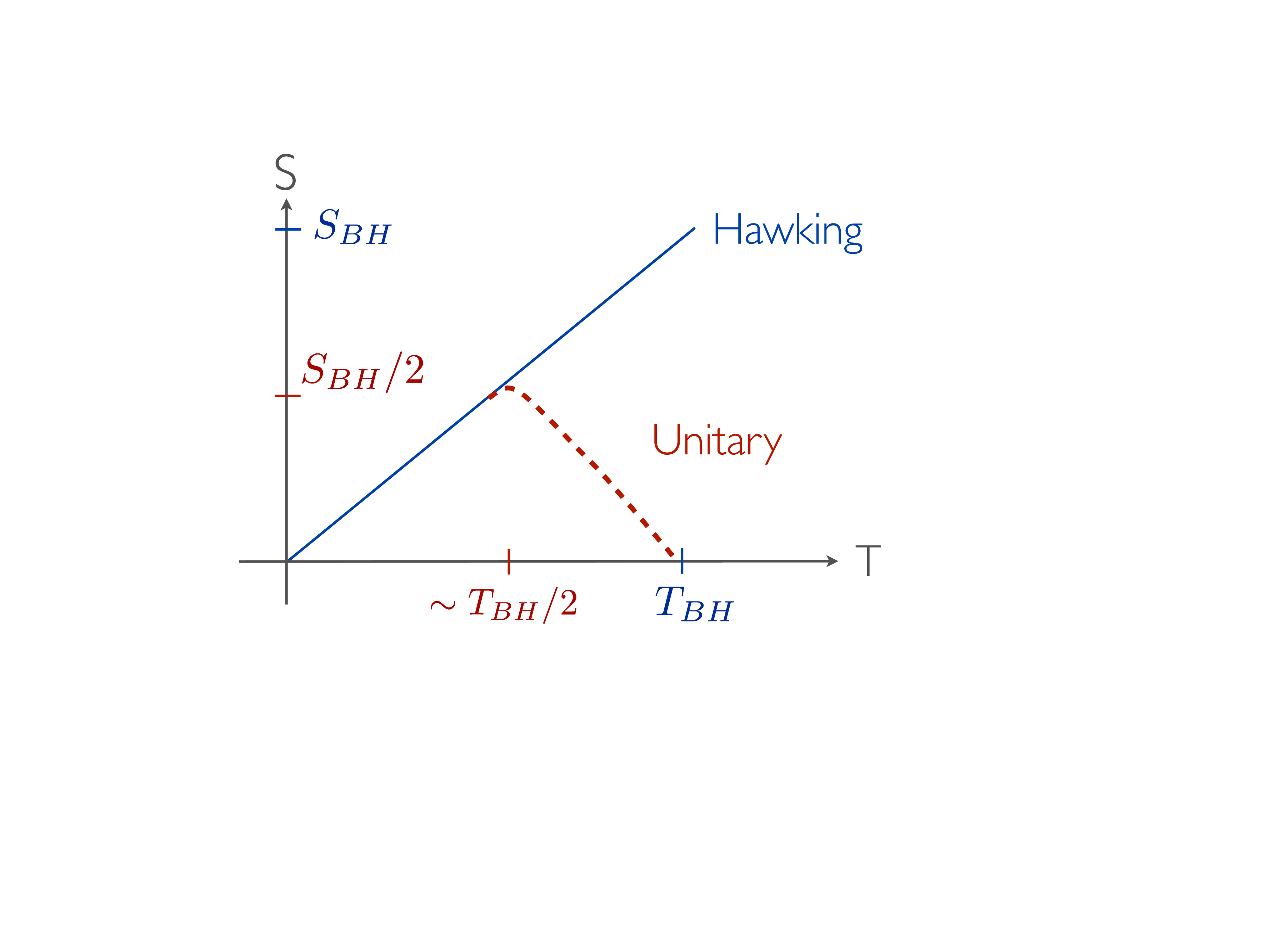}{3.5}

The von Neumann entropy of the Hawking state, \vN, describes the information missing from the external state.  This grows with time as sketched in \entropfig. If the black hole disappears completely at the end of this evaporation, the final entropy, $S(T_{BH})\sim S_{BH}$, quantifies the non-unitarity of the evolution.  While unitary evolution would be possible if this information remained in a black hole remnant, that scenario also appears ruled out (see {\it e.g.}  \refs{ \wabhip, \Sussrem}).  

For unitary evolution, the curve $S(T)$ therefore must drop back to $S=0$ at the end of evolution, also as sketched in \entropfig.  Indeed, Page\refs{\Pageinfo} has argued that, under certain assumptions, the curve turns over when the entropy of the radiation matches that of the black hole, roughly at the black hole half-life.

While it has been argued in \refs{\QBHB} (see also \GiddingsSJ) that the nice slice description does not sharply approximate the correct $S(T)$ beyond $T\sim \calo(R S_{BH})$, which would resolve the actual ``paradox," an essential question is what mechanics leads to unitary evolution.  Such evolution seems not to be described by local quantum field theory.

If we assume quantum mechanics governs nature, we expect a Hilbert-space description of the physics.  While dynamics may not be exactly local, the discussion of section two suggests that a possibly ``coarser" notion of locality remains, in decompositions of the total Hilbert space into factors corresponding to subsystems.

The preceding section has outlined such a decomposition, in the case where the physics is LQFT.  Since this apparently violates unitarity, we seek a modification to the physics, beyond LQFT.  But, we will assume that it has a more general quantum-mechanical description, based on Hilbert space and unitary maps, and has localization properties arising from tensor-factor structures.

To be concrete, let us recapitulate  the Hilbert-space structure and evolution arising from LQFT in the previous section, and then examine how it might be modified.  As outlined there, the space of physical states can be described as lying in a larger product of Hilbert spaces, $ \hbh(T)\otimes\hext(T)$, which can be thought of as corresponding to modes that are inside or outside the black hole at time $T$ in some slicing.  

\subsubsec{States}

In particular, LQFT together with the ``nice" slicing provides a model for $\hbh$: it is described by excitations of modes, either ``outgoing" or ingoing, that impact the constant-$r$ part of the nice slice.  As they do so, they freeze, due to the vanishing lapse.  This in fact provides essentially the expected state counting.  The dominant modes of the Hawking radiation are those with wavelength $\sim R$ and one of these is typically emitted each time interval  $R$.  The total number of ``partner" modes hitting the internal slice during the evaporation time of the black hole is thus $\sim S_{BH}$.  There can also be a contribution from ingoing modes.  To penetrate the black hole these should have wavelength $\roughly< R$.  There are many modes at shorter wavelength that can fall into the black hole, but if we count incoming states that do not significantly perturb the black hole's evolution, this requires the total energy of infalling matter be less than the initial mass of the black hole.  The dominant states saturating this condition are again $\sim S_{BH}$ particles of energy $\sim 1/R$.  

We might expect that the external Hilbert space is well-described by the familiar Fock space construction; indeed, whatever the light degrees of freedom are, we expect them to behave asymptotically like propagating particle states (and hence all states in a Minkowski background to ultimately have such a labeling).
For purposes of studying modifications to LQFT structure, though, it is helpful to make a further distinction, between modes of $\calh_{ext}(T)$ that are ``near" the black hole, and ``far,"  
\eqn\outdec{\calh_{ext}(T)=\calh_{near}(T)\otimes\calh_{far}(T)\ .}
Specifically, modes whose central distance, measured in the spatial slice geometry,  is larger than, say, $5R$ might comprise the latter, and the rest, the former ``atmosphere"  of the BH.

Then, the general state lies in a product
\eqn\bnfdecomp{\calh=\hbh(T)\otimes\hnear(T)\otimes\hfar(T)\ ,} 
on which the unitary LQFT evolution operator \unitop\ acts.

We seek ``minimal" modifications to LQFT that result in unitary evolution, assuming that the basic structure \bnfdecomp\ of the Hilbert space, where states lie in a product space corresponding to different ``localizations at a given time," is present in the {\it full} theory of quantum gravity.  In this context $T$ is a parameter which might be identified with asymptotic time.  If LQFT is minimally modified, a natural expectation is that the ultimate description of $\hfar(T)$ is essentially unmodified, for sparse populations of low-energy asymptotic particles; such states should have a good LQFT description.  As will be discussed, the space $\hnear(T)$ may or may not be modified, depending on assumptions.  For $\hbh$, we expect significant modification.  In particular, the missing information in Hawking evolution results from the final dimension of $\hbh$ being of order $\exp\{S_{BH}(M_0)\}$ for a black hole of initial mass $M_0$.   Instead, in unitary evolution we expect this dimension to shrink to one by the end of evolution, so no information is contained.  Indeed, an obvious ansatz for the dimension is 
\eqn\bhdim{{\cal N}(M) \approx \exp\{S_{BH}(M)\}\ ,}
with $M$ the value of the mass at a given time $T$.

\subsubsec{Evolution: basic features}

We can also consider plausible forms for evolution, with minimal modification of LQFT.  Again, we expect essentially unmodified evolution of the form \unitop\ on $\hfar$.  Likewise, one might expect that transport of modes from $\hnear$ to $\hfar$ or the reverse is essentially governed by LQFT evolution \unitop, unless there are large modifications to $\hnear$.  But, there must be significant modifications to the unitary evolution describing interaction of the black hole interior $\hbh$ with the black hole atmosphere $\hnear$, since this evolution must transport the information initially in $\hbh$ into $\hnear$, and this is forbidden by locality.   Even so, we can seek to describe $\hbh$ and its evolution in a fashion with ``least" departure from LQFT.  Such evolution is relevant for describing infalling observers, and one naturally expects them to be governed by LQFT until their demise at the center of the black hole.

In short, the problem becomes one of describing unitary evolution on the subsystems $\hbh$, $\hnear$, and $\hfar$.  During this evolution, the dimensions of the individual subsystems may change, {\it e.g.} as in \bhdim, but a generalized notion of unitarity (one-one, linear, inner-product preserving maps) remains.

One can then seek to constrain such evolution.  One reasonable set of constraints is the match to LQFT dynamics described for $\hext$.  Other constraints will further tighten the description.

\subsec{Quantum  information transfer: general constraints}

In quantum theory, all information is fundamentally quantum information.  The present problem is to investigate transfer of this information between the subsystems $\hbh$ and $\hext$, also allowing for change in their dimensions.   For practical purposes, we expect these can be taken as finite dimensional.  For $\hbh$ this is an essential part of the story of how unitarity is recovered -- the alternative leads to missing information or remnants.  In the case of $\hext$, this can also be thought of as a good approximation, since the whole system can be regarded as contained in a very large box, and one considers only sufficiently low-energy states in that box.

So, we have a basic problem in quantum information theory, which is to characterize and constrain (generalized) unitary maps of the form
\eqn\transf{U:\ \ha\otimes \hb \rightarrow \ha'\otimes \hb'}
which transfer information from system A (here BH) to system B (here ext).  Some general features of this will be described here, with further development in \refs{\GiSh}.

A basic question is what constitutes information transfer.  An approach to describing this is via a trick used in \HaPr. Namely, introduce an auxiliary space $\hc$ that is a copy of $\ha$.  Given a basis $|I\rangle$ for A, with dimension $\caln_A$, a density matrix with maximal entanglement between A and C arises from the state\foot{Note that in the case of Hawking evolution and the Hawking partners in $\hbh$, the external particles function much as $\hc$, as seen from \statepack.}
\eqn\acstate{|\chi\rangle = {1\over \sqrt{\caln_A}} \sum_I |I\rangle_A |I\rangle_C\ .}
Specifically, in this state the von Neumann entropy \vN\ of $\rho_A={\rm Tr}_C(|\chi\rangle\langle\chi|)$ is $S_A=\log\caln_A$, and likewise for $\rho_C={\rm Tr}_A(|\chi\rangle\langle\chi|)$.  The unitary evolution \transf\ is extended to $\ha\otimes \hb\otimes \hc$ as 
\eqn\unitext{U\otimes 1_C\ .}
(Again, compare Hawking evolution.) Consider a state $|\psi\rangle=|\chi\rangle |\phi \rangle$, for some $|\phi\rangle\in\hb$.  Under evolution \unitext, the entropy of $\rho_{AB}={\rm Tr}_C(U|\psi\rangle\langle\psi|U^\dagger)$ remains constant at $S_{AB}=S_C=\log\caln_A$.  In effect the auxiliary system C is used to ``tag" the information.  Specifically, with $\rho_A={\rm Tr}_{BC}(U|\psi\rangle\langle\psi|U^\dagger)$, decrease of $S_A$ corresponds to information transfer ``out of A."  At the same time, the entropy $S_B$ of $\rho_B={\rm Tr}_{AB}(|\psi\rangle\langle\psi|)$ will increase.  A general constraint is subadditivity,
\eqn\subadd{S_{A}+S_B\geq S_{AB}\ .}

In particular a unitary map \transf\ reducing the dimension of A reduces $S_A$.  By \subadd, $S_B$ will then increase.  Increase such that the subadditivity inequality is saturated corresponds to a definite kind of ``minimal" information transfer.

In fact, saturation of subadditivity implies\GiSh\ that the unitary map transfers information in a particularly simple way: modulo unitary transformations acting on the subsystems A and B, the transformation transfers a $k$-dimensional subspace from A to B.  Suppose $\ha$ is a tensor product $\calh_k\otimes\calh_{\caln_A/k}$, with product basis $|i\rangle|a\rangle$, and let $|b\rangle$ be a basis for $\hb$.  Then, such a minimal unitary transformation of the form \transf\ may be defined by its action on the bases,
\eqn\subtrans{(|i\rangle|a\rangle) |b\rangle \rightarrow |a\rangle (|b\rangle |i\rangle)\ .}
where $|a\rangle$ gives a basis for $\ha'$ and $|b\rangle |i\rangle$ for $\hb'$.  A transformation of this particular form, modulo unitary transformations acting on the individual subsystems, will be called {\it subspace transfer}.  A special case of such transfer, for $k=2$, is qubit transfer.

Non-saturating transfer is non-minimal in the sense that there is extra ``excitation" of the B subsystem for a given amount of information removed from A\GiSh.  In fact, a particularly simple example of a non-saturating transformation is
\eqn\nmex{ |0\rangle_A |0\rangle_B\rightarrow |0\rangle_{A'} |0\rangle_{B'}\ ;\  |1\rangle_A |0\rangle_B\rightarrow |1\rangle_{A'} |1\rangle_{B'}\ .}
 Here the B subsystem is initially one-dimensional (hence trivial), and the dimension of the product Hilbert space grows by a factor of two.  In this simple case information is not transferred out of A, but correlations are developed with B.

Given the relative simplicity of subspace transfer, a first question in characterizing candidate unitary evolution laws in the black hole context is whether they saturate subadditivity.  For example, evolution posited in \refs{\Pageinfo,\PageDF} is saturating.  In addressing this question, we turn to other expected physical constraints on the evolution.

\subsec{Physical constraints and characterization}

In order to further constrain evolution, let us consider possible physical constraints on a family of transformations of the form \transf, describing black hole evolution, with $\ha=\hbh$ and $\hb=\hext$.  Some of these constraints will be essential; others, while plausible, may not be universally agreed upon.

A first constraint which we regard as essential is:

\item{A.} {\it The final entropy of the system, evolved from an initially pure state, is zero: evolution is unitary.}

Another basic constraint is that  there is an asymptotic notion of energy (at least for asymptotically Minkowski systems, with possible generalizations), and 

\item{B.} {\it Energy is conserved.}

A constraint that many consider plausible is:

\item{C.} {\it The evolution should appear innocuous to an infalling observer crossing the horizon; in this sense the horizon is preserved.}

There are other possible constraints.  One is

\item{D.} {\it Information escapes the black hole at a rate $dS/dt\sim 1/R$.  }

Another possible expectation is that the radiation remains ``Hawking-like"  in other respects.  One characteristic of this is 

\item{E.} {\it The coarse-grained features of the outgoing radiation are still well-approximated as thermal.}

Additionally, in line with the above, a possible constraint is

\item{F.} {\it Evolution saturates the subadditivity inequality \subadd.}

Finally, evolution given by the operators $U$ needs to be part of a complete, consistent framework.  And, this framework should have the basic property of correspondence: in situations outside of the extremes of black holes or other strongly gravitational situations, the rules should approximately reduce to those of LQFT together with semiclassical general relativity, to a good approximation

It is not a priori clear that there {\it are} such evolution laws satisfying even the most important of these constraints, particularly consistency, correspondence, and A-C: here, specifically, we encounter the conflict with LQFT.  For this reason, it seems worth investigating {\it any} evolution laws that do satisfy such basic constraints.\foot{Note also that the constraints may turn out not to be independent.}  An additional plausible guideline in this is that we should be as conservative as possible, and seek to describe evolution that is ``as close as possible" to that of LQFT.

\subsec{Unitary models: large departures from LQFT}

\subsubsec{Fast scrambling}

Let us begin with a candidate form of evolution that appears not to satisfy the ``most-conservative" dictum.  Suppose an ingoing mode falls into the black hole, or consider the partner mode to a Hawking particle.  In the nice-slice, LQFT model for $\hbh$ and the dynamical $U_{LQFT}$, we have seen that the mode freezes at $R_c$, thus adding another factor to the Hilbert space $\hbh$.  A modification of this is to consider action  on $\hbh$ of a unitary ${\hat U}$ that is essentially random, followed by transfer of a particle (described as subspace transfer) to outgoing radiation in $\hext$.  If $|i\rangle$ denotes the state of the ingoing mode, and $|\ahat\rangle$, $|b\rangle$ bases for the rest of the states of $\hbh$ and $\hext$, the first two steps are of the form
\eqn\fastone{|\ahat\rangle |i b\rangle \rightarrow |\ahat i\rangle |b\rangle\rightarrow ({\hat U} |\ahat i\rangle) |b\rangle\ ,}
followed by subspace transfer in a new basis:
\eqn\fastwo{|\ahat' i'\rangle |b\rangle\rightarrow |\ahat'\rangle | i'b\rangle\ .}
(Here, we simplify notation from \subtrans.)  We refer to the transformation given by ${\hat U}$ as ``scrambling;" we might moreover assume it acts on a time-scale $T_{sc}\sim R$ or $\sim R\log R$ \refs{\HaPr,\SeSu,\Sufast} which is ``fast."  A sequence of such scrambling transformations, together with sufficient subspace transfer to continuously reduce dim$\hbh$, by imprinting information in outgoing states, can clearly accomplish the ultimate $S\rightarrow0$.

Such a modification is clearly a large departure from LQFT nice-slice evolution; to quantify this, the scrambling time for Hawking evolution on nice slices is $T_{sc}=\infty$.  Another way to describe the large departure is provided by Hayden and Preskill\HaPr.  They show that if the internal black hole space evolves via such scrambling, then after sufficient time, additional information thrown into the black hole will accessible in the external state on the scrambling timescale -- black holes would behave as information {\it mirrors}.  

Indeed, one quantitative characterization of candidate evolution laws is such a scrambling time.  While Hawking evolution predicts $T_{sc}=\infty$,  we have reviewed arguments\refs{\QBHB,\GiddingsSJ,\GiSl} that nice-slice evolution is not a good description past the timescale $\tbh\sim R S_{BH}$; also, Page's arguments regarding ``generic" (though subadditivity-saturating) evolution point to a similar timescale.  Scrambling faster than $T_{sc}\sim R\log R$ would produce a contradiction\refs{\STU}; {\it at} this  timescale evolution may minimally satisfy condition C, and permit an infalling description for time $R\log R$, though the associated statements of horizon complementarity appear at odds with condition C.\foot{For example, horizon complementarity posits that observables inside and outside the black hole are complementary variables analogously to $x$ and $p$ in quantum mechanics.  Note that such a picture may correspond to a particular choice of gauge, like that given by the Schwarzschild slicing.  If so, a more germane question is whether there is a gauge, for example like those based on nice or natural slices, permitting both an inside and outside description.  For further general discussion of gauge transformations, see sec.~five.}

 But there is a large range of  timescales between $R\log R$ and $R S_{BH}$.  And, the latter  timescale for information return, if part of a consistent picture, is ``more conservative" in that it is closer to the value given by LQFT.

\subsubsec{Massive remnants/fuzzballs}

 \Ifig{\Fig\MRevol}{Sketch of possible evolution of a massive remnant or fuzzball, in the Eddington-Finkelstein geometry of \EddFink.  An infalling observer following the arrow is expected to be disrupted upon impacting the surface.}{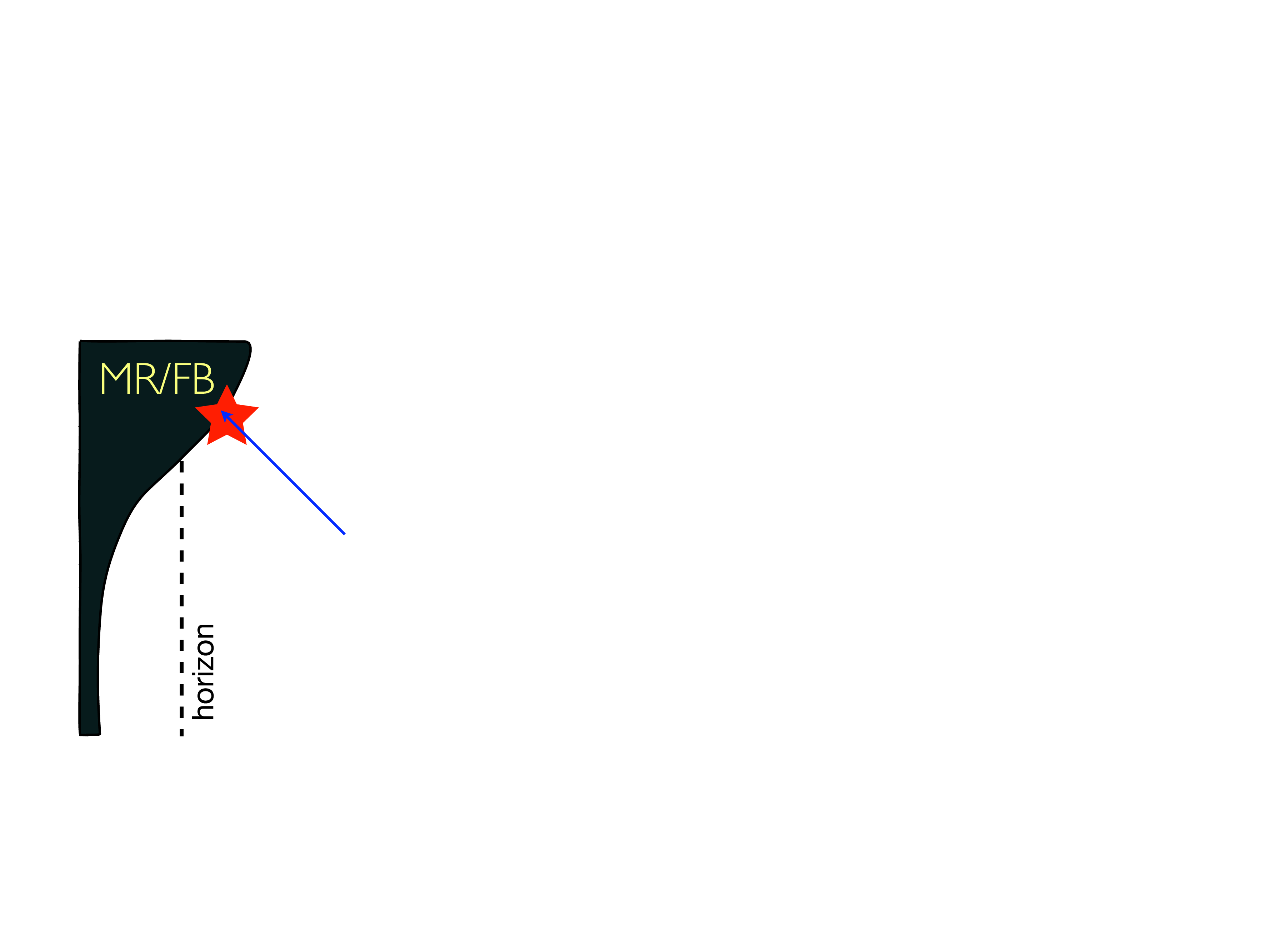}{2}

Another apparently less-conservative general scenario is that of massive remnants\refs{\BHMR}, or what seems to be a recent realization of it, the fuzzball scenario\refs{\fuzz}.  The basic picture here is that, due to unknown dynamics, the information inside the black hole ultimately expands to give an object with significant modification to the semiclassical geometry outside the would-be horizon, perhaps as sketched in \MRevol.  Propagation of the surface of such a remnant would be outside the light-cone, and in that sense non-local.  

This scenario thus violates condition C.  A big modification of the geometry outside the horizon generically has a big effect on infalling observers.  An example with some possible similarity is a neutron star -- infalling observers rapidly scramble with the neutrons near their impact point, though not immediately with the entire star.

The fuzzball scenario, if it can be realized for highly non-BPS objects like Schwarzschild black holes, would appear to fit into this category.  The reason is that this scenario is commonly described as accounting for the large information with a large class of geometries with significant departure from that of the black hole, outside the horizon.  A superposition of such geometries would appear to have a rapidly varying microstructure, and thus be very disruptive to infalling matter.\foot{One should note, however, attempts to avoid this conclusion\refs{\MaPl}.}

Other apparently large departures from semiclassical evolution, {\it e.g.} \tHooftAA, might also be described in a similar fashion. Different evolution with significant deviation from LQFT appears in \Rozali.

\subsec{Unitary models: minimal departure from LQFT?}

One is naturally led to ask whether there are candidates for consistent evolution that are closer to that of LQFT.  This subsection will explore a class of such candidates.

We again assume that at a given stage in the evolution the total Hilbert space takes the product form \bnfdecomp.  Here $T$ is a parameter, which we can identify with asymptotic time.  Evolution is given by unitary maps (in the sense of \transf) $U(T',T)$ that map \bnfdecomp\ into the analogous Hilbert space at time $T'$.

\subsubsec{Description of $\hnear$ and $\hfar$}

As suggested in sec.~4.1, we do not expect significant modifications to a LQFT description of $\hfar$.  Likewise, we assume that the part of $U(T,T')$ that acts on $\hfar$ is well-approximated by the LQFT expression \unitop.  We can also try to stay close to LQFT by positing that the structure of $\hnear$ is largely unmodified, as is propagation \unitop\ between $\hnear$ and $\hfar$.  (The preceding evolution laws depart from this to different degrees.)    

A refinement of this is to posit that the {\it state} in $\hnear$ does not have large departures from the Hawking state.
This can be a significant constraint, since then there are few ``active" modes present in $\hnear$.  Specifically, in terms of the localized bases for outgoing modes described in sec.~3.4, few modes are relevant.  Modes with $\omega_j\gg 1/R$ are exponentially suppressed by thermal factors, as are high occupation numbers.  Modes with $l\gg \omega_j R$ have essentially no effect on the outside dynamics, as LQFT largely forbids their transport to $\hfar$.  Finally, the cutoff $a<A(T)$ and the limitation to the near regime $r\roughly<5R$ limits the range of $a$.

Indeed, there is a further limitation, arising from a strong form of condition C.  If one modifies the Hawking state, for example by adding or subtracting a quantum in a mode with short wavelength as seen on the slice, that will be seen by an infalling observer as a high-energy near-horizon excitation.\foot{For example, such alteration spoils the cancellation \refs{\GiddingsSJ} of interactions with infalling quanta.}   If we require that the infalling observer see no excitations above an energy not much larger than $1/R$, this then restricts the range of $a$ for which there can be significant modification of the Hawking state.  Thus, for wavepacket resolution parameter $\epsilon\sim 1/R$, just a few values of $a$, depending on the cutoff energy, are active.  Another way to say this is that the Hawking state typically corresponds to emission of one particle of wavelength $\sim R$, in the s-wave, per time $R$, and we only permit alteration of such quanta when they have evolved  to near this asymptotic wavelength.  The nearby active modes in $\hnear$ are thus limited to a small number.

States satisfying this restriction can  be characterized in an occupation number basis $|\{n_{jal}\}\rangle$ based on a set of modes like the ${v}_{ja}$ of \vki.  We first require
\eqn\hcons{H=\sum_{jal} n_{jal} \omega_j < K \beta^{-1}\ ,}
where $K$ is some finite number, say $K<5$.  This restricts the allowed occupied $\omega_j$'s, as well as the occupation numbers.  The relevant range of $l$ is restricted by the condition $l<K \omega_j R$ (though inessential, for simplicity the same $K$ is used).  The allowed range of $a$, for a given $T$, is (again simplifying)
\eqn\arange{ {\epsilon\over 2\pi}(T-KR) < a< {\epsilon\over 2\pi}(T+KR)\ .}
We will streamline the notation by relabeling the basis for such states, {\it e.g.} $|\{n_{jal}\}\rangle$, simply as  $|I\rangle$.
  This construction provides a model for the outgoing states of $\hnear(T)$, which has a finite and moderate-size dimension, with  $\sim K^2/(2\pi)$ modes (independent of $\epsilon$) for small $\epsilon$.  Ingoing states can also be included, and the two kinds can mix under \unitop\ due to reflection.

\subsubsec{$\hbh$}

As outlined in section 3.4, LQFT also provides a model for the $\sim\exp\{S_{BH}\}$ states of a black hole, for example as quantum field theory excitations on a nice slice with internal length $\sim R S_{BH}$.  However, such an $\hbh$ does not reduce its dimension through LQFT evolution, resulting in the unitarity crisis.  Thus, one expects that significant modification is needed both to the LQFT model of $\hbh$ and to its evolution.  This expectation is reinforced by arguments that perturbative LQFT evolution does not describe the black hole state past a time $\sim R S_{BH}$.  

In particular, we expect that $\hbh$ can be modeled as a Hilbert space of finite dimension $\caln(M)$, with $M$ the BH mass, and that this decreases past a certain point in BH evolution, possibly as in \bhdim.  At the same time, unitary evolution requires that the information be transferred to $\hnear$, as outlined in sec.~4.1.

\subsubsec{Evolution}

Thus, the essential problem is to describe unitary evolution acting on the finite dimensional spaces 
\eqn\hbhcomb{U(T',T): \hbh(T)\otimes \hnear(T)\rightarrow  \hbh(T')\otimes \hnear(T')\ ,}
such that the dimension of the first factor decreases to zero at the end of evolution, to satisfy condition A, above.  If we view LQFT as giving a good description of the black hole interior, at least for a limited time, we may also model at least the most-recently infallen modes by the LQFT description.  However, by a time $\sim RS_{BH}$ such a description needs to be significantly modified.\foot{In descriptions corresponding to other slicings, such as the natural slices, the description may be modified even sooner.}  In addition, $U_{LQFT}$ acts to transport modes between $\hnear$ and $\hfar$, and to evolve modes in $\hfar$.  

There are various characteristizations of this evolution.  First, unitary evolution can act to mix modes within $\hbh$, as described in the discussion of fast scrambling.  As noted, $T_{sc}=\infty$ for Hawking evolution in the nice slicing (neglecting backreaction).  Second, we need unitary evolution to transfer information from $\hbh$ to $\hnear$.  This process may have a different associated  timescale $T_{tr}$.  Following the discussion of section 4.2, this transfer may be minimal (in that it saturates subadditivity), or non-minimal.  Given that the transfer is expected to be a weak process, and also in the case of  small dimension of $\hnear$, which restricts the deviation from saturation\GiSh, we will focus on minimal transfer, and leave non-minimal possibilities for later investigation.

To describe evolution, let $|\hat a\rangle$ give a basis for $\hbh(T)$.  Then, nice-slice Hawking evolution for an interval $\Delta T=T'-T$ of order the light-crossing time for $\hnear$ takes the basic form (see \statepack)
\eqn\hawkU{|\hat a\rangle\rightarrow |\hat a\rangle\sum_I e^{-\beta H/2}|{\hat I}\rangle|I\rangle\ ,}
together with both outward and inward evolution of other modes via $U_{LQFT}$.  Here, $|{\hat I}\rangle$ represents a copy of $\hnear$ corresponding to the Hawking ``partners" we have described.

\subsubsec{Models}

The modifications we seek reduce the dimension of $\hbh$.  One set of models retains the Hawking partners, to stay close to LQFT, but modifies the evolution:
\eqn\hawkmod{|\hat a\rangle\rightarrow U(\Delta T)(|\hat a\rangle |{\hat I}\rangle|I\rangle)\in \hbh(T')\otimes \hnear(T')\ .}
Such evolution is parameterized by finite-dimensional matrices
\eqn\hawkmodt{U(\Delta T)(|\hat a\rangle |{\hat I}\rangle|I\rangle) = U_{\ahat {\hat I} I}^{{\hat a}'{I}'} |{\hat a}'\rangle |{ I}'\rangle  \ ;}
we also expect that for small $\Delta T$, $U$ is close to unity and so may be parameterized in ``hamiltonian" form.

Simple models for such evolution can be given by describing the states in terms of qubits\refs{\HaPr,\Mathurrev,\ModelsU}, and within these models some examples of evolution were outlined in \ModelsU.  (Related models appear in \refs{\AveryNB}.)  Modeling the information as contained in one such qubit mode with frequency $\omega$, one is
\eqn\evoltwo{\eqalign{\zhats\zhats \ahats \as &\rightarrow \uhat \ahats\otimes \left(\zhats \zs+e^{-\beta\omega/2} \ohats\os\right)\otimes U \as\ ,\cr  \zhats\ohats \ahats \as &\rightarrow \uhat \ahats\otimes\zhats\os\otimes U\as\ ,\cr \ohats \zhats \ahats \as &\rightarrow \uhat \ahats\otimes\ohats\zs\otimes U\as\ ,\cr \ohats\ohats \ahats \as &\rightarrow \uhat \ahats\otimes \left(e^{-\beta\omega/2}\zhats \zs- \ohats\os\right)\otimes U \as\  }} 
up to trivial normalizations of the states. Here, we have split off a subspace corresponding to the ``first" two bits of $\hbh$, and information from them is transferred into $\hnear$.  We also allow for unitaries $U$, ${\hat U}$ acting individually on $\hbh$ and $\hnear$, whether or not given by LQFT evolution. 

This may be extended to a more realistic model as follows.  Suppose that when a black hole forms, most of the degrees of freedom of $\hbh$ are in a fiducial ``vacuum" state, in accord with the statement that collapsing matter produces far fewer than $\exp\{S_{BH}\}$ black hole states.  In the qubit model, this could be described by most qubits being in state $\zhats$.  The small factor of $\hbh$ corresponding to excited degrees of freedom is associated with states of the infalling matter -- $\ohats$ in the qubit model.  Then, unitary evolution acts both to scramble the states on  timescale $T_{sc}$ and transfer the inside information out on  timescale $T_{tr}$.  The transfer might work as in \evoltwo, namely if an unexcited ($|{\hat 0}\rangle$) degree of freedom gets transferred out, it maps to the Hawking state of ${\widehat{\cal H}}_{\rm near}\otimes \hnear$, but excited degrees of freedom in $\hbh$ produce other, orthogonal, states.  Even a rapid scrambling time for $\hat U$ and short $T_{tr}$ is then expected to produce close to the Hawking state, for a very long time, $\sim R S_{BH}$.  Or, an alternative is that scrambling time is much longer (even $> RS_{BH}$), or that scrambling acts on a subset of $\hbh$, for example the first $S_{BH}/10$ bits.  Of course $T_{tr}$ can be long, but is bounded as $\roughly< R S_{BH}$ for each bit; ultimately one needs information tranfer rates $\roughly> 1/R$.  

One expectation is that the generic such models produce an extra flux, beyond the (approximately) thermodynamic flux predicted by Hawking.  This can happen since generically the information transfer, \hawkmodt,  can populate modes that are not excited in the Hawking state, adding to its outward flux.  One example\ModelsU\ illustrating this phenomenon can be described in qubit language as 
\eqn\evolthree{ |\qhat_1\qhat_2\rangle \ahats \as \rightarrow \uhat \ahats\otimes \left(\zhats \zs+ e^{-\beta \omega/2} \ohats\os\right)\otimes|\zhat'\zhat''\rangle|q_1'q_2''\rangle\otimes U \as\ ,}
In this example information from internal degrees of freedom (here  the first two qubits) is transferred into modes of $\hnear$ (here $|q_1'q_2''\rangle$) that are not typically populated in the Hawking state.  So, in such a model a black hole disintegrates faster than predicted by Hawking, once information transfer begins to be important.  (This still may happen only on timescale as long as $\calo(RS_{BH})$.)

We might refer to evolution \hbhcomb\ which yield thermodynamic fluxes like the Hawking state as ``strongly Hawking-like."  An interesting question is what assumptions imply such evolution, as opposed to more rapid disintigration, if  the latter is present in more complete models. For example, in the preceding evolution, $E$ decreases faster than implied by $dE=T dS$, unless there is a modification to the temperature, which is determined by the density of states. One possible way to achieve thermodynamic behavior is through a version of  the evolution described below \evoltwo, where the scrambling acts on a large number of degrees of freedom -- say for example the ``first half" of the $\sim S_{BH}$ degrees of freedom on a nice slice -- leading to effective thermalization.  Then, the thermal distribution might be transferred out by evolution of the form \evoltwo.

This discussion has only given a preliminary view of unitary evolution models of the general form \hbhcomb.  Additional criteria and constraints (for example those discussed in sec.~4.3) can be employed to refine understanding of such models, for example in the parameterization \hawkmodt.    Investigation of the resulting constraints on both information scrambling and transfer and on the broader dynamics are left for future work.

\subsubsec{Nonlocality, and effect on infalling observers}

Evolution laws like \hbhcomb, that transfer information from the internal states of a black hole to its ``atmosphere," represent a departure from a local description with respect to the semiclassical BH geometry, and thus an apparent departure from the framework of LQFT.  It should be borne in mind that, in the perspective explored in this paper, the semiclassical  geometry is not necessarily fundamental to the physical description, so such nonlocality with respect to this geometry could simply represent a shortcoming of a picture based on this spacetime geometry.  To take a parallel from quantum mechanics, this picture might be as incorrect as description of quantum particle motion in terms of sharp classical trajectories in phase space.

Whether such nonlocality is well-phrased with respect to the geometry, or rather represents a shortcoming of that description, such apparent nonlocality offers another seeming advantage.  Specifically, one puzzle in any attempt to describe information escape from a black hole is how to respect condition C, that this appear innocuous to infalling observers.

In particular, as has been noted, if the information were transferred to modes when they are at the horizon, and of very short wavelength, that would be seen by the infalling observer as a large perturbation to the vacuum:  he/she would see high-energy particles at the horizon, and sufficiently many of these would even destroy the horizon.  But, if transferral of information is taking place in a fashion that is nonlocal with respect to the semiclassical geometry, there is no clear reason to insist that the information transfer just be to modes at the horizon -- one might equally permit transferral to modes in the atmosphere region of size $\sim R$ surrounding the horizon, described by $\hnear$.  

Such transferral represents a disruption of the Hawking state, but one that can be harmless.  In particular, roughly one bit needs to be transferred to the atmosphere per time $R$, if the black hole starts radiating information near its half-life.  In the preceding description, this alters a small number of modes -- of order one per time $R$ -- seen to be of energy $\sim 1/R$ by the infalling observer.  For a large black hole, such transfer rates to such low-energy quanta seems completely innocuous.  So, the single assumption that locality is modified apparently can avoid this kind of potential issue.

One other potential concern is that information transferral outside the lightcone could lead to causality paradoxes.  In Minkowski space, spacelike communication can be converted into communication into the past, by a Lorentz transformation.  However, if the present phenomenon only arises in certain strongly-gravitating contexts, this is not necessarily an issue.  In particular, Lorentz boosts are not a symmetry of the Schwarzschild spacetime.  So, one cannot obviously convert such ``spacelike communication" into acausal propagation -- the overall picture can apparently remain causal\refs{\NLvC}.

\newsec{Towards a general framework: possible outlines of a nonlocal mechanics}

Ultimately it is essential to have a clearer picture of an overarching framework describing the kinematics and mechanics of  such a theory with a modified notion of locality.  This paper explores the viewpoint that {\it the basic structure underlying physics is Hilbert space, not spacetime.}  A first question is how to formulate quantum mechanics sufficiently generally to describe physics in a situation where space and time are not necessarily part of the fundamental description, but are ``emergent."  This in particular means that one should not formulate physics in terms of sums over spacetime histories, as with  generalized quantum mechanics\refs{\HartGQM}.  Some basic postulates for a more general formulation of quantum mechanics are given in \refs{\UQM}.   

Of course, more structure needs to be added to a Hilbert-space description in order to approximately recover spacetime and the dynamics of quantum fields in that spacetime.  As described above, one can see a possibly more fundamental origin of the notion of locality, or more precisely localization, in factorization of a Hilbert space into tensor factors describing different ``regions."  Then, the information that can be recovered about the resulting approximate geometry should be encoded in relationships between the factors; factors can be nested, like open sets in a geometry, or overlapping, producing a factor that corresponds to the intersection (\Tensfact).  
Spacetime structure should thus be approximately reconstructed by the ``net" of interlocking tensor factors, and additional spacetime structure is not necessarily input at the beginning.  (Here, again, is one important difference from the approaches of algebraic quantum field theory\Haag, and holographic spacetime\BanksHST, which associate observable algebras or Hilbert spaces with pre-existing causal spacetime diamonds.)  For describing both this structure, and the spectrum of particles, one does need additional information about how the Hilbert spaces are labeled, and interconnected.

Another important aspect of locality arises in evolution, namely locality limits how states in different regions, here tensor factors, can interact and communicate.  In the present approach to black holes, the conflict between locality and unitarity is assumed decisively resolved in favor of the former: unitarity rules.  Nonetheless, we expect to be able to describe approximately local QFT evolution in a wide range of contexts, and in particular all that are familiar to present experiment.  Correspondingly, one expects limitations on the unitary evolution.  Specifically, localization structure enters through Hilbert-space factorization, and evolution should appropriately respect this localization structure.
Indeed, locality in LQFT states that signals don't propagate outside the light cone.  A central point of the current work is that this LQFT notion of locality needs to be modified in the strongly gravitating context, but is expected to hold in other contexts.  In Hilbert space language, it might be described in terms of the time required for evolution  to cause information to traverse ``across" a given tensor factor.  But more work is needed to understand constraints on such evolution, and in particular under what circumstances one recovers LQFT evolution in a correspondence limit.  We have just begun the exploration of possible evolution -- though in a nontrivial context where the clash between locality and unitarity is manifest.

Symmetry also plays an essential role in physics.  For example, if $\calh$ is a Hilbert space corresponding to a spacetime with an asymptotic symmetry group ({\it e.g.} Minkowski space, anti-de Sitter space), then, in accord with Wigner's theorem, one expects a unitary operator $\cals: \calh\rightarrow\calh$ implementing each such symmetry -- consider for example an overall boost of the system.\foot{Lorentz symmetry thus requires an infinite-dimensional Hilbert space corresponding to the arbitrarily large ``center-of-mass" momentum.}  Then, the dynamics should also respect the symmetry: $\cals^\dagger U\cals = U$.  

If locality is modified, a more nontrivial question is how local symmetries are approximately recovered.  If, for example, $\calh=\calh_1\otimes \calh_2\otimes\cdots$, we expect a global  $\cals$ to arise from individual actions on the factors:
\eqn\symfact{\cals \calh=\cals \calh_1\otimes \cals \calh_2\otimes\cdots\ .}
For example,  picture an overall translation or boost acting on states in spatially separated regions -- states in each region should undergo the same translation or boost.
Then, a local transformation can arise by such transformations acting differently in different regions: 
\eqn\sloc{\cals_{\rm loc}\calh = \cals_1\calh_1\otimes \cals_2 \calh_2\cdots\ .}

Indeed, we see such a structure in the LQFT limit.  Consider for example \unitop.  We can translate the ``left" half of spacetime, $x<0$, forward in time, but leave the ``right" half, $x>0$, fixed, by acting with such a $U_{LQFT}$ with $N$ that vanishes on the right but not on the left.  This corresponds to a change of slicing, $t'=t'(t,x^i)$, $x^{i\prime}=x^i$; by combining such transformations one realizes ``multi-fingered time."  Such a transformation is of the general form \sloc, and in particular on small enough factors of the Hilbert space acts simply as a time translation.  One can likewise consider local spatial translations, or boosts or rotations, to obtain more general diffeomorphisms.

Thus, it seems reasonable to hypothesize that in a more basic Hilbert tensor network description of the dynamics, the same kind of structure realizes a version of local symmetry transformations appropriate to the localization structure.  Transformations of the form \sloc, which in field theory terms relate different choices of slicings (and spatial coordinatizations), can be thought of more generally as relating different descriptions of the space of states.\foot{From this perspective, we might see limitations of the nice-slice description of the black hole state as arising from limitations on the kinds of extreme gauge transformations necessary to go to the nice slice gauge for long time spans, particularly $\sim RS_{BH}$.  By such timescales, matching of the Hilbert spaces may be badly distorted from that of LQFT.}  One expects other local symmetries ({\it e.g.} internal gauge symmetries) could be similarly realized.

In spacetime, implementing diffeomorphism symmetry as a symmetry of the evolution requires introduction of a metric of non-Minkowski form, as one sees in the LQFT limit through the entrance of the metric in \hgen, \unitop.  So, in such a Hilbert-space framework one expects further information about the nature of gravity from realizations of the corresponding symmetry.  The just-noted connection with the equivalence principle -- that  actions on small tensor factors are expected to be ``local" translations or boosts -- is expected to guide description of the dynamics of matter in a gravitational background, as with general relativity.  But -- also as with general relativity -- the question of  determining the dynamics of gravity seems a more challenging question.  

Indeed, formulating a complete theory seems a daunting challenge, but we have many constraints described here, and elsewhere (for example based on properties of the S-matrix\Erice).  Again, if the present situation is similar to the development of quantum mechanics, we can seek encouragement in the fact that, once headed in the right direction, there were two paths to the correct physics, via matrix and wave mechanics.  Or, recall that LQFT is essentially determined from the very general assumptions of quantum mechanics, locality, Poincare symmetry (and other symmetries), and the existence of particles.  So, if we are indeed headed in the right direction, possibly, once again, the rigidity of structure surrounding correct physics will provide crucial guidance.

\bigskip\bigskip\centerline{{\bf Acknowledgments}}\nobreak

I thank T. Banks, R. Emparan, J. Hartle, M. Hastings, P. Hayden, D. Marolf, D. Mateos, S. Mathur, D. Morrison,  Y. Shi, M. Srednicki, and W. van Dam for useful conversations.  This work  was supported in part by the Department of Energy under Contract DE-FG02-91ER40618 and by grant FQXi-RFP3-1008 from the Foundational Questions Institute (FQXi)/Silicon Valley Community Foundation.

\listrefs
\end